\documentclass[10pt,lettersize,journal]{IEEEtran}
\ifCLASSINFOpdf

\else

\fi
\usepackage{amssymb,amsmath,epsfig}
\makeatletter
\def\maketag@@@#1{\hbox{\m@th\normalfont\normalsize#1}}
\makeatother
\usepackage{graphicx}
\usepackage{epstopdf}
\usepackage{microtype}
\usepackage{cite}
\usepackage{url}
\usepackage[none]{hyphenat}
\usepackage{booktabs}
\usepackage[linesnumbered,ruled,vlined]{algorithm2e}
\usepackage{algorithmicx}
\usepackage{algpseudocode}
\usepackage{multirow}
\usepackage{nicefrac}
\everymath{\displaystyle}
\usepackage{color}
\usepackage[font=footnotesize]{caption}
\usepackage[linesnumbered,ruled,vlined]{algorithm2e}
\usepackage{algorithmicx}
\graphicspath{{Schematics/}} 
\usepackage{algpseudocode}
\usepackage{url}
\usepackage{hyperref}
\usepackage{dblfloatfix}  
\usepackage{subcaption}
\usepackage{graphicx}
\usepackage[table,xcdraw]{xcolor}
\DeclareMathOperator*{\argmin}{argmin} 
\begin{document}
	%
	\title{{Index Modulation-based Information Harvesting for Far-Field RF Power Transfer}}

		\author{M. Ertug Pihtili~\IEEEmembership{Student Member,~IEEE}, Mehmet C. Ilter~\IEEEmembership{Senior Member,~IEEE},  Ertugrul Basar~\IEEEmembership{Fellow,~IEEE},\\  Risto Wichman~\IEEEmembership{Member,~IEEE}, and Jyri  H\"am\"al\"ainen~\IEEEmembership{Senior Member,~IEEE}
		\thanks{This work was supported by the Academy of Finland (grant number: 334000).  The preliminary versions of this paper have been published in Proc. IEEE Glob. Commun. Conf. (GLOBECOM), 2022 [26] and Proc. IEEE Veh. Tech. Conf. (VTC-Spring), 2023 [27]. 
        \par M. Ertug Pihtili and E. Basar are with the Communications Research and Innovation Laboratory (CoreLab), Department of Electrical and Electronics Engineering, Koc¸ University, Sariyer, Istanbul 34450, Turkey. (e-mail: mpihtili22@ku.edu.tr; ebasar@ku.edu.tr)
        \par Mehmet C. Ilter, Risto Wichman, and Jyri  H\"am\"al\"ainen are with the Department of Information and Communications Engineering, Aalto University, Espoo, Finland (e-mail: mehmet.ilter@aalto.fi; risto.wichman@aalto.fi; jyri.hamalainen@aalto.fi).}}	
	
	\IEEEtitleabstractindextext{%
		\begin{abstract}
			While wireless information transmission (WIT) is evolving into its sixth generation (6G), maintaining terminal operations that rely on limited battery capacities has become one of the most paramount challenges for Internet-of-Things (IoT) platforms. In this respect, there exists a growing interest in energy harvesting technology from ambient resources, and wireless power transfer (WPT) can be the key solution towards enabling battery-less infrastructures referred to as zero-power communication technology. Indeed, eclectic integration approaches between WPT and WIT mechanisms are becoming a vital necessity to limit the need for replacing batteries. Beyond the conventional separation between data and power components of the emitted waveforms, as in simultaneous wireless information and power transfer (SWIPT) mechanisms, a novel protocol referred to as information harvesting (IH) has recently emerged. IH leverages existing WPT mechanisms for data communication by incorporating index modulation (IM) techniques on top of the existing far-field power transfer mechanism. In this paper, a unified framework for the IM-based IH mechanisms has been presented where the feasibility of various IM techniques are evaluated based on different performance metrics. The presented results demonstrate the substantial potential to enable data communication within existing far-field WPT systems, particularly in the context of next-generation IoT wireless networks. 
		\end{abstract}
		
		}
	
	\maketitle
\IEEEdisplaynontitleabstractindextext
	
\IEEEpeerreviewmaketitle

\section{Introduction}

\IEEEPARstart{I}{n} {the 6G era, IoT platforms are required to support a massive number of devices and to tackle energy demands in addition to existing challenges in data transmission.} In this respect, existing discussions in standardization frameworks are underway in ambient IoT scenarios \cite{Zhang2022}, where IoT devices are able to harvest ambient energy from external sources. {In this direction, zero-power communication technology \cite{Naser2023} has been emerged as the next logical step in future networks and promises information transmission where its energy is harvested from surrounding radio frequency energy, thus reducing the need for replacing batteries\cite{Liu2022}. To reach  greater sustainability and environmentally friendly architectures, the synergistic integration of various frameworks connecting energy harvesting with communication infrastructure appears to be an inevitable development in the near future.}

The origin of WPT can be traced back to the seminal work of N. Tesla \cite{tesla1904transmission}. His idea was based on radiating the energy through a medium (i.e., air) with the help of antenna elements. Nowadays, the WPT has been extensively studied for radio frequency (RF) signals \cite{Valenta2014, Gu2021} and for visible light communications \cite{Pan2019}, which is an emerging 6G paradigm. Particularly, the WPT can be categorized into two main classes: near-field and far-field. The former applies power transfer over short distances by magnetic fields using inductive coupling or electric fields between the energy harvester and the power transmitter, which is out of the scope of this paper. The latter utilizes ambient RF waves in the surrounding indoor/outdoor environment resulting from existing RF transmission along with backscattering techniques and provides an ideal green power source for longer ranges \cite{Clerckx2019}.

In this regard, the far-field power transfer introduces viable solutions to tackle the battery depletion problem for wireless nodes/devices. Energy harvesting enables wireless nodes to scavenge energy from the environment without requiring battery replacement and tethering to electricity grids \cite{Ku2016}. The ongoing surge in interest prevails across a broad spectrum of applications, which are remote environmental monitoring, consumer electronics, biomedical implants, building/home automation, Industry 4.0, and logistics solutions \cite{Leemput2023}. In the far-field power transfer, the primary emphasis has centered on enhancing RF-to-DC conversion efficiency where the rectennas, which combine rectifiers and antennas, are key elements in the energy harvester. Consequently, the development of efficient rectennas was a longstanding focus in earlier literature \cite{Clerckx2019}. These investigations have highlighted the significance of tailoring rectenna designs to match precise operating frequencies and input power levels, thus presenting a significant challenge due to the inherent nonlinear characteristics in practical rectenna implementations \cite{Costanzo2016}. 

Furthermore,  the efficiency of RF energy harvesting also depends on the choice of a selected WPT waveform at the power transmitter. For instance, it was shown that deploying a multi-sine waveform increases the efficiency of RF-to-DC conversion, so the output DC power \cite{shen2021} for flat-fading channels. Aside from grid-based and lattice-based constellations used in conventional communication systems where low peak-to-average-power-ratio (PAPR) is typically desired due to nonlinearity in power amplifiers, real Gaussian signals, flash signaling, and linear frequency-modulated signals are preferred in earlier far-field power harvesting mechanisms \cite{Clerckx2019} {at the cost of higher complexity and power consumption in the transmitter due to their higher  PAPR.} Interestingly, the effects of different modulation techniques, such as amplitude shift keying (ASK), quadrature amplitude modulation (QAM), phase-shift keying (PSK), and frequency shift keying (FSK) on battery charging time, were measured and theoretically analyzed in \cite{Cansiz2020}.
 
	
	
 Wireless powered communication networks (WPCN) were initially proposed for reducing the operational workload of battery replacement/recharging over RF-based energy harvesting systems \cite{Suzhi2016}. The network elements in WPCN first harvest energy from the signals transmitted by RF energy sources and then consume this harvested energy for their upcoming communication periods. Rather than only radiating energy via RF signaling, incorporating data communication into wireless power transfer emerged during the last decade. This is mainly referred to as simultaneous wireless information and power transfer (SWIPT) over RF-based mechanisms \cite{Ponnimbaduge2018} and, more recently, as simultaneous light information and power transfer (SLIPT) for optical-based ones have been introduced \cite{Uysal2021}.  In those systems, the power and information components are mostly separable from each other over different domains, which can be the power domain (power splitting), time domain (time splitting), and space domain (antenna splitting) \cite{Liu2016}.  From this aspect, there exists a trade-off between information transfer and energy transfer based on the design preferences of the data and power transmission. There is an extensive body of work exploring these preferences within the trade-off \cite{Amarasuriya2016}, and SWIPT studies can also be found in commercial RFID systems, particularly in the context of communication from reader to RFID tags \cite{Paolini2022}. Additionally, the SWIPT mechanism has been extended into multi-user cases, referred to as multi-user (MU)-SWIPT, where a multi-antenna transmitter simultaneously transmits wireless information and energy through spatial multiplexing to multiple single-antenna receivers~\cite{Zhang2014}.
 
In order to secure ongoing data transmission in conjunction with a long-range wireless power transfer mechanism, it should be kept in mind that most devices are simple nodes, i.e., RedCap devices \cite{Veedu2022} at the level of several kbps, having limited computational capabilities, thus sending information and power simultaneously makes the SWIPT mechanism disadvantageous in practice due to different device capability requirements. The results presented in \cite{Xu2019} demonstrate that power transfer efficiency also drops dramatically over the transmission distance in such systems when security enhancement was added to the design. To overcome this constraint, a distributed antenna-based SWIPT protocol was proposed in \cite{Huang2018}, and \cite{Krikidis2019}, where information bits are embedded in the tone index of multi-sine waveforms. Alternatively, an approach involving the creation of a modulation signal within a vacant resource block of communication in an orthogonal frequency division multiplexing (OFDM) block has also been introduced in \cite{Nakamoto2022}.

\begin{figure}[!t]
    \centering
    \includegraphics[width=0.78\columnwidth]{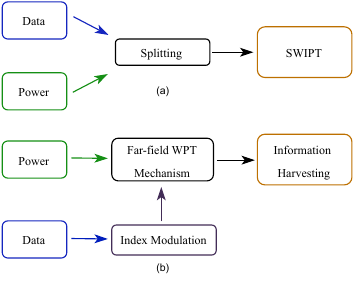}
    \caption{The design aspects of (a) SWIPT  and (b) Information Harvesting \cite{Ilter2022}  mechanisms which show how data and power parts are integrated together.}
    \label{fig:IHnutshell}
\end{figure}

Thus, there is a need for complementary solutions to combat this potential interception
of confidential information at unintended users. Although conventional encryption techniques are useful in many cases for protecting data transmission, there are some use cases/protocols in which additional encryption cannot be applied \cite{Wang2019}. In this respect, artificial noise (AN)-based physical layer security (PLS) techniques  \cite{Goel2008} turned into a powerful tool due to its ability to generate orthogonal noise through the channel state information between the transmitter and intended receiver.  Considering its advantages against eavesdropping, the AN technique has been applied to multiple-input single-output (MISO) \cite{Lv2018} and multiple-input multiple-output (MIMO) systems \cite{Gu2019}.

To address the power transfer efficiency losses and implementation concerns arising from varying sensitivities between the energy harvester and the data recipient mentioned above, an alternative approach has been proposed in \cite{Ilter2022}, allowing for information transfer alongside existing wireless power transfer without disrupting the ongoing WPT mechanism between the power transmitter and the energy harvester. This approach is referred to as \textit{Information Harvesting (IH)}, and \figurename{ \ref{fig:IHnutshell}} illustrates the differences in the integration of data and power elements in these scenarios compared with the SWIPT scenarios. As seen in the figure, the SWIPT employs a splitting mechanism, while IH focuses on harnessing the available WPT mechanism. This approach enables information transfer through WPT without compromising the range of the wireless power transfer service area, a common limitation in SWIPT systems. This is accomplished by encoding information into the indices of transmitter entities through the implementation of index modulation (IM) techniques. {The generalized space shift keying (GSSK)-based IH mechanism and its performance in terms of secrecy capacity and average harvested power was first investigated in \cite{Ilter2022GL}.}  Then, \cite{Ilter2023} introduced a  new  IH  mechanism that relies on quadrature spatial shift keying (QSSK).

Another superiority of the IM-based IH lies in its reduced pilot overhead since the MU-SWIPT system necessitates the channel state information (CSI) of both the information receiver (IR) and the energy harvester (EH) links. In contrast, the IH systems do not require CSI of the EH link, potentially leading to a reduced pilot overhead for channel estimation. Additionally, the IH mechanism distributes the total power among the active transmit antennas, a strategy arising from embedding information within the transmit entities of the system. This distinctive feature positions the IH mechanism as a promising solution, expanding the service area of the far-field WPT mechanism without excluding simple nodes, especially IoT devices in need of energy. 

In this paper, we introduce the IH architecture within a unified and comprehensive framework that encompasses a broader spectrum of IM techniques. We investigate the feasibility of these techniques while establishing unified performance criteria. In this aspect, our contributions can be summarized as follows:
\begin{itemize}
    \item Currently, the  IH mechanisms were limited to space shift keying-based mechanisms and lacked of presenting a wider perspective. In this paper, IM-based IH mechanisms are presented in a comprehensive manner where not only space shift keying-based techniques with no modulated symbol but also spatial modulation ones have been included. 
    \item Unlike only concentrating on certain performance metrics, this paper introduces a wider scope for performance evaluation where the different IM-based IH mechanism has been investigated through energy harvesting capability, bit error rates, and ergodic secrecy rates. In addition, a unified framework for the calculation of the theoretical error performance of the proposed schemes and provide useful insights. The simulated error rate results are validated by analytical error rate expressions derived for the proposed IM-based IH mechanisms.
    \item The superiority against existing multi-user SWIPT mechanism \cite{Zhang2014} has been presented along with other practical advantages of the IM-based IH mechanisms.
\end{itemize}

The remainder of this paper is organized as follows: In Section II, we present the general structure of IM-based IH model where the different implementations in the power transmitter based on the choice of IM technique, how the energy harvester and information receiver operate based on the chosen IM technique are introduced. The unified error performance analysis is given in Section III where the average bit error rate in the information receiver and the eavesdropper is derived, respectively.  Section IV presents the ergodic capacity analysis in a similar manner, and the feasibility of the IM-based IH mechanism is investigated in Section V via a variety of simulation results along with the validation from the analysis. Section VI completes the paper with concluding remarks. For clarity, the abbreviations and notations in the text are listed in Table~\ref{tab:Abbreviations}.

\begin{table}[]
\captionsetup{justification=centering, labelsep=newline}
\caption{\textsc{ABBREVIATIONS AND NOTATIONS.} \label{tab:Abbreviations}}
\centering
\footnotesize
\resizebox{0.9\columnwidth}{!}{%
\begin{tabular}{|c|l|}
\hline
\textbf{Abbreviation} & \textbf{Meaning}                                                \\ \hline
6G                   & Sixth generation                                                \\ \hline
AN                   & Artificial noise                                                \\ \hline
ASK                  & Amplitude shift keying                                          \\ \hline
AWGN                  & Additive white Gaussian noise                                  \\ \hline
BER                  & Bit error rate                                                  \\ \hline
CPEP                 & Conditional pairwise error probability                          \\ \hline
CSI                  & Channel state information                                       \\ \hline
CSCG                 & Circularly symmetric complex Gaussian                           \\ \hline
CW                   & Continous wave                                                  \\ \hline
DC                   & Direct current                                                  \\ \hline
EH                   & Energy harvester                                                \\ \hline
EIRP                  & Effective Isotropic Radiated Power                             \\ \hline
ESR                  & Ergodic secrecy rate                                            \\ \hline
FSK                  & Frequency shift keying                                          \\ \hline
GSM                  & Generalized spatial modulation                                  \\ \hline
GSSK                 & Generalized space shift keying                                  \\ \hline
GQSM                 & Generalized quadrature spatial modulation                       \\ \hline
GQSSK                & Generalized quadrature-based space shift keying                 \\ \hline
IH                   & Information harvesting/harvester                                \\ \hline
IR                   & Information receiver                               \\ \hline
IoT                  & Internet of Things                                              \\ \hline
IM                   & Index modulation                                                \\ \hline
ML                   & Maximum likelihood                                              \\ \hline
MU-SWIPT             & Multi-user simultaneous wireless information and power transfer \\ \hline
OFDM                 & Orthogonal frequency division multiplexing                      \\ \hline
PAPR                 & Peak to average power ratio                                     \\ \hline
PDF                  & Probability density function                                    \\ \hline
PLS                  & Physical layer security                                   \\ \hline
QAM                  & Quadrature amplitude modulation                                 \\ \hline
QSM                  & Quadrature spatial modulation                                   \\ \hline
QSSK                 & Quadrature-based space shift keying                             \\ \hline
RF                   & Radio frequency                                                 \\ \hline
RFI                 & Request for information                                          \\ \hline
RFID                 & Radio frequency identification                                  \\ \hline
SLIPT                & Simultaneous light-wave information and power transfer           \\ \hline
SM                   & Spatial modulation                                              \\ \hline
SSK                  & Space shift keying                                              \\ \hline
SVD                  & Singular value decomposition                                    \\ \hline
SWIPT                & Simultaneous wireless information and power transfer            \\ \hline
WIT                  & Wireless information transfer                                   \\ \hline
WPCN                 & Wireless powered communication networks                         \\ \hline
WPT                  & Wireless power transfer                                         \\ \hline
WPTx                  & Wireless power transmitter                                         \\ \hline
\hline
\textbf{Notation}                            & \textbf{Definition}                                                                                                                                  \\ \hline
$x$                                          & Scalar values                                                                                                                                        \\ \hline
$\mathbf{x}$, $\mathbf{X}$                   & Vectors/Matrices                                                                                                                                     \\ \hline
$\left(\cdot\right)^T$                       & Transpose operation                                                                                                                                  \\ \hline
$\left(\cdot\right)^H$                       & Hermitian transpose                                                                                                                                  \\ \hline
$|\cdot|$                                    & Absolute value                                                                                                                                       \\ \hline
rank(.)                                      & Rank of matrix                                                                                                                                       \\ 
\hline
$\log_a (\cdot)$                                  & Logarithm with base $a$                                                                                                                              \\ \hline
$||\cdot||$                                  & Norm of vector/matrix                                                                                                                                \\ \hline
$\mathbb{E}[\,\cdot\,]$                      & Average operator                                                                                                                                     \\ \hline
$\mathbb{E}_\mathbf{x}[\,\cdot\,]$           & Statistical expectation with respect to random variable $x$                                                                                          \\ \hline
$P\left(\cdot\right)$                        & Probability of an event                                                                                                                              \\ \hline
$\binom{n}{r}$                               & Binomial coefficient                                                                                                                                \\ \hline
$\lfloor\cdot\rfloor$                        & Floor operation                                                                                                                                     \\ \hline
$M$                                          & Modulation order                                                                                                                                     \\ \hline
$N$                                          & Number of subbands in WPT waveform                                                                                                                                     \\ \hline
$p\left(\cdot\right)$                        & Probability density function                                                                                                                         \\ \hline
$\mathcal{Q}(x)$                             & Q-function                                                                                                                                           \\ \hline
$j$                                          & Imaginary unit, $\sqrt{-1}$                                                                                                                          \\ \hline
$\Re\{\cdot\}$                               & Real part                                                                                                                                            \\ \hline
$\Im\{\cdot\}$                               & Imaginary part                                                                                                                                       \\ \hline
$\mathcal{CN}\left ( \mu, \sigma^2 \right )$ &  CSCG random variable with a mean $\mu$ and a variance $\sigma ^2$ \\ \hline
$\mathbb{C}^{m \times n}$                    & A set of complex matrices of $m \times n$ dimensions                                                                                                 \\ \hline
\end{tabular}%
}
\end{table}
\section{IM-based Information Harvesting Model}
\begin{figure}[t]
		\centering
		\includegraphics[width=0.9\columnwidth]{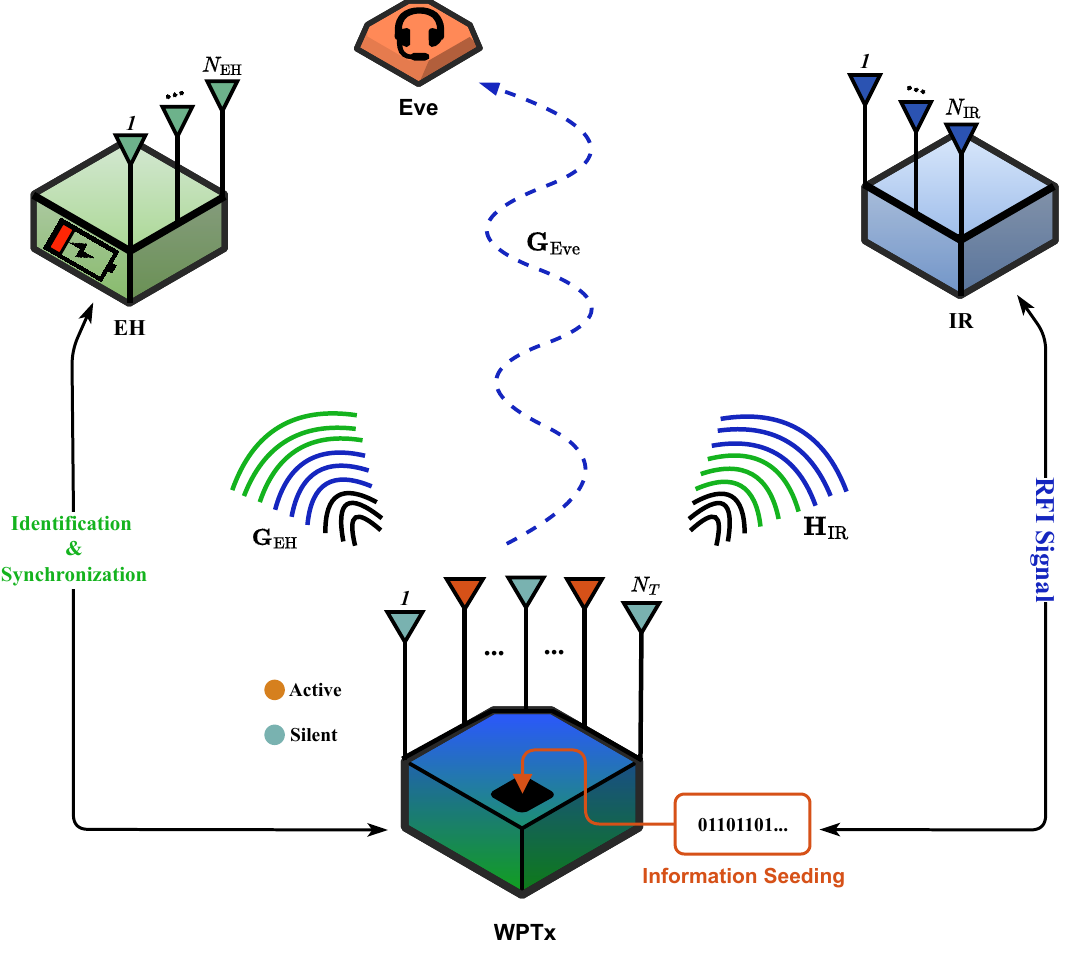}
		\caption{The block diagram of the IM-based IH mechanism where the WPTx serves EH while sending its information to the IR by utilizing IM techniques with the existence of Eve in a service area.}
		\label{fig:IH_Sys}
\end{figure}

The block diagram of the IM-based IH mechanism is illustrated in Fig.~\ref{fig:IH_Sys}. In this configuration, it is assumed that the energy harvester (EH) and information receiver (IR) are located within the service area of the wireless power transmitter (WPTx) node, along with a potential malicious node, Eve. Initially, WPTx, comprising a total of $N_{\rm T}$ transmit antennas, serves as the node responsible for transmitting multitone WPT waveforms, which utilize $N$ distinct subbands, into the area for RF energy harvesting. It is assumed that identification and synchronization between WPTx and an EH device are successfully established. When an IR device enters the service area, WPTx aims to transmit available information blocks to the IR. This is achieved by mapping the information blocks into vectors corresponding to active transmit antennas, a process called \textit{information seeding.} Meanwhile, the EH device, equipped with $N_{\rm EH}$ receive antennas, conducts recharging operations without any disruption. Notably, when the IR device enters the service area, there already exists a far-field WPT mechanism facilitated by a wireless link denoted as $\mathbf{G}_{\rm EH}\in\mathbb{C}^{N_{\text{EH}} \times N_T \times N }$.

Subsequently, upon deployment, the IR device, which is equipped with $N_{\rm IR}$ receive antennas, initiates an information seeding cycle. This initiation is performed by transmitting a request-for-information (RFI) signal over a dedicated link represented by $\mathbf{H}_{\rm IR}\in\mathbb{C}^{N_{\text{IR}} \times N_T \times N}$. The RFI signal may also include configuration parameters on the information seeding cycle, such as the frequency of changing the transmitting entity based on the required data rate within the IR and the necessity for additional PLS protocols \cite{Ilter2022}. The information seeding process is activated upon detecting the RFI signal at the WPTx. During this phase, the IR aims to capture variations resulting from the activation of various transmit antenna sets, which are determined by the information bits. This process is referred to as \textit{information harvesting.} Importantly, it was demonstrated in \cite{Ilter2022} that activating more transmit antennas in WPTx during the power transfer period reduces the probability of detecting information seeding activity by devices located even in close proximity to the EH. 
Note that the channels between the WPTx and other network elements in Fig.\ref{fig:IH_Sys} are assumed to be stationary, so the time dependency of the channel coefficients is emitted from the channel coefficient and weight factors in the rest of the paper.
\subsection{EH: Energy harvesting}

During the period when the WPTx emits power transfer waveforms into a service area (regardless of the activation of information seeding cycle), the EH converts the received RF signal into DC output power thanks to its rectennas. Particularly, the EH consists of a receive antenna chain along with $N_{\rm EH}$ receive antennas, battery charging unit, and battery. Herein, the battery charging unit is configured to establish a link between the receive antenna chain and battery, wherein the manner in which power is transferred from the wireless power transceiver is controlled according to the parameters and/or state information assigned by the power management unit \cite{Ilter2023}. 

{The EH device operates in two stages: firstly, the received RF power is converted into DC power using rectennas. Note that due to the nature of the wireless medium, there exists a fraction of time where the rectenna cannot perform harvesting since input RF power lies below certain RF power that is called as\textit{ rectenna sensitivity} and after certain received signal power level at harvester, \textit{rectenna saturation power}, the harvested energy stays constant as shown in \cite{Alevizos2018}. After considering these practicalities as in \cite{Clerckx2022}}, the output DC power for rectenna $q$ can be formulated as a function of the received RF signal, which is,
	\begin{equation}
		{
			{v}_{out, q}=\left \{ \begin{aligned}
				0&, \quad\quad P_r^t \in \left[0, \Gamma_{in} \right]  \\ 
				\beta_2 P_r^t  + \beta_4 {P_r^t}^2 &, \quad {P_r^t} \in \left[\Gamma_{in}, \Gamma_{sat} \right] \\ 
				\beta_2 \delta_{sat}  + \beta_4 \delta_{sat}^2 &, \quad {P_r^t} \in \left[\Gamma_{in}, \Gamma_{sat} \right]
			\end{aligned}  \right.}
		\label{eq:EHharvesting}
	\end{equation}
{where $P_r^t$ is the received input power {during coherence period $t$}, $P_r^t~=~\mathbb{E}[|y\left(t\right)|^2]$,  $\Gamma_{in}$ refers the harvester sensitivity, $\Gamma_{sat}$ denotes the saturation level, $\beta_2$ and $\beta_4$ are the  parameters of the nonlinear rectifier model.}  Then, the outputs of all rectennas are aggregated to obtain the cumulative harvested DC power at the combiner's output. This technique is referred to as  DC combining, as discussed in \cite{Clerckx2022}. Then, the total DC output power can be obtained from $\textstyle \sum_{q=1}^{N_{\rm EH}}{v}_{out,q}^2/R_L$ where $R_L$ refers to the resistive load used to determine the output DC power.

{In the nonlinear rectenna model, the calculation of the harvested DC power is not straightforward. For this purpose, \cite{Bayguzina2016} introduced a new technique where Taylor series expansion was applied into the nonlinear diode model. This approach provides insight into how the choices of the signal modulation or the input distributions affect the energy harvesting capability along with the configuration of the rectenna. Inspired from it, in this paper, the energy harvesting capability relates with a  variable of $z_{DC}$ which establishes a direct connection with \eqref{eq:EHharvesting} and reflects the harvested power when the signal operates within the bounds of the linear and saturation regions. Mathematically, $z_{DC}$ is formulated as \cite{Bayguzina2016} }

\begin{equation}
    z_{DC} = k_2R_{ant}\mathbb{E}[|y(t)|^2] + k_4R_{ant}^2\mathbb{E}[|y(t)|^4].
    \label{eq:Zdc}
\end{equation}

In a steady-state response, an ideal rectifier maintains a constant output voltage over time, and the level of this output voltage depends on the peaks of the input voltage. Therefore, $z_{DC}$ can be improved as the number of subbands increases in the emitted waveform from the WPTx. This enhancement is attributed to the generation of larger peaks facilitated by multitone signals, despite their average power being equivalent to that of continuous wave (CW) signals. The impact of multitone signals, which results in the generation of larger peaks, leads to a higher PAPR. Consequently, higher PAPR enhances the efficiency of the RF-to-DC conversion process in the nonlinear rectifier, contributing to an improved energy harvesting capability. This enhancement is particularly associated with the incorporation of a greater number of subbands in the emitted waveform. In this respect,  \eqref{eq:Zdc} demonstrates a positive correlation with PAPR, suggesting that modulation schemes or input distributions characterized by higher PAPR values offer advantages to the nonlinear rectenna model \cite{Kim2020}.

\subsection{WPTx: Information seeding}
In this subsection, we discuss the integration of different IM schemes into IH mechanism along with detailed description of the information seeding cycle. IM represents a unique approach to information transmission, achieved by selectively activating specific elements for conveying information.  When information seeding is initiated in WPTx alongside the specified IM scheme, the resulting set of waveforms emitted from the $N_a$ activated WPTx antennas, obtained by combining an IM waveform at the WPTx, is as follows:
\begin{equation}
\mathbf{s} = \left[{\frac{s_1(t)}{\sqrt{N_{a}}}}, 0, \dots, 0, \frac{s_m(t)}{\sqrt{N_{a}}}, 0, \dots, 0, \frac{s_{N_{a}}(t)}{\sqrt{N_{a}}} \right], \label{eq:TransmitSignal} \end{equation}
at a given time instant $t$. Hereby, $s_m(t)$ is directly associated with the selection of the WPT waveform, $m \in \{1, \dots, N_T\}$. For instance, when employing a multitone signal with $N$ distinct subbands, the expression for $s_m(t)$ can be formulated as:
\begin{equation}
	s_m(t)=\Re\Big\{\sum_{n=1}^{N}\omega_{n,m}e^{j 2\pi f_n t}\Big\}.
	\label{eq:mthantennatransmit}
\end{equation}

{
\noindent Hereby, $N$ represents the number of subbands in the generation of multitone WPT signals and $f_n$ denotes the center frequency of the $n$th subband waveform. More specifically, $\omega_{n,m}~=~ a_{n,m} e^{j\theta_{n,m}}$ where $a_{n,m}$ and $\theta_{n,m}$ are the $n$th subband amplitude and phase components of baseband  WPT signal and these values are based on the chosen IM scheme. Subsequently, the received passband signal at the IR, without the additive white Gaussian noise (AWGN) term, can be mathematically expressed as given by \cite{Bayguzina2016},
}
\begin{equation}
\mathbf{y}_{\text{IR}}(t)=\Re \Big\{ \sum_{n=1}^{N}\textbf{G}_{\text{EH,n}} \textbf{w}_n e^{j 2\pi f_n t}\Big\},
	\label{eq:receivedsignal}
\end{equation}
{where $\textbf{G}_{\text{EH},n}$ is the channel between WPTx and IR for $n$th subband and $\mathbf{w}_n = [\omega_{n,1} \dots \omega_{n,N_T}]^T$, where the power of $\mathbf{w}_n$, denoted as $\lambda_s^2\slash N $ (i.e., $\mathbb{E}[\mathbf{w}_n^H\mathbf{w}_n] = \lambda_s^2\slash N$ ), is arranged in accordance with the values of $N$ and $N_a$.}


\subsubsection{GSSK/SSK-based IH}
In the GSSK-based IH mechanism, available information bits are only mapped into a transmit antenna vector, which determines the number of active transmit antennas, denoted as $N_{\rm a}$, with the constraint $N_{\rm a} \leq N_{\rm T}$ \cite{Jeganathan2008}, when $N_{\rm a} = 1$, it simplifies to the SSK-based IH mechanism. The number of information bits that can be mapped into the transmit antenna indices for GSSK modulation is given by:
\begin{equation}
	\eta_{GSSK} =\left \lfloor \log_2\left(\begin{matrix}N_{T} \\ N_{a} \end{matrix} \right) \right \rfloor,
	\label{eq:informationbits}
\end{equation}
where $\lfloor \cdot \rfloor$ indicates floor operation. Under this assumption, the emitted WPT waveform does not contain any information components itself; instead, the antennas simply emit power transfer waveforms to the service area. From the EH perspective, there is no difference in the GSSK/SSK-based IH mechanism, as no signal modulation is employed, leading to $a_{n,m} = 1$ and $\theta_{n,m}= 0$. Note that $L=2^{\eta}$ different active antenna combinations exist when $N_{a}$ antennas are active out of total $N_T$ transmit antennas, which is equal to the cardinality of GSSK codebook.  



\subsubsection{GSM/SM-based IH} In the case of GSM-based IH, WPTx begins embedding information within the emitted symbol. This differs from the SSK-based techniques mentioned earlier, where only unmodulated WPT waveforms exist. Consequently, the number of transmitted information bits can be expressed as:
\begin{equation}
	\eta_{GSM} =\left \lfloor \log_2\left(\begin{matrix}N_{T} \\ N_{a} \end{matrix} \right) \right \rfloor + \log_2 \left(M\right),
	\label{eq:informationbits}
\end{equation}
where $M$ represents the modulation order of $M$-QAM constellation used in the WPTx. In the GSM-based IH mechanism, $a_{n,m}$ and $\theta_{n,m}$ should be configured according to the employed QAM. For instance, in case of 4-QAM is utilized for GSM/SM-based IH mechanism, the $a_{n,m}$ and $\theta_{n,m}$ values can be set to $\{\sqrt{2}$, $7\pi\slash4\}$ for a bit sequence of $\left[0 1\right]$ in addition to the bits mapped into active antenna indices.

\subsubsection{GQSSK/QSSK-based \hspace{-0.37cm}IH} Quadrature-based implementations require additional signal processing before transmission to the information seeding stage. The transmitted symbols are decomposed into in-phase and quadrature components, resulting in two transmit antenna vectors. The first transmit antenna vector corresponds to the real part of a wireless power symbol, while the second corresponds to the imaginary part \cite{Mesleh2015}. Therefore the number of information bits increases to
\begin{equation}
	\eta_{GQSSK} =2\eta_{GSSK}.
	\label{eq:informationbits2}
\end{equation}
{Assuming $s_m\left(t \right)$ is a complex WPT signal, and its the real part, $\Re\{s_m\left(t \right)\}$,  and its imaginary part, $\Im\{s_m\left(t \right)\}$, are transmitted separately through the different antennas over cosine and sine carriers respectively \cite{Mesleh2015}.} In the GQSKK/QSKK-based IH mechanism, the transmitted symbol is fixed and equal to $1+j$. Consequently, the corresponding $a_{n,m}$ and $\theta_{n,m}$ values are $\{\sqrt{2}, \pi/4\}$.

\subsubsection{GQSM/QSM-based IH}
Similar to \eqref{eq:informationbits}, now WPT signal carries the modulated symbol from $M$-ary constellation, and the real part modulates the in-phase part of the carrier, whereas the imaginary part modulates the quadrature component of the carrier signal. Then, the spectral efficiency of GQSM is  given by
\begin{equation}
	\eta_{GQSM} =2 \left \lfloor \log_2\left(\begin{matrix}N_{T} \\ N_{a} \end{matrix} \right) \right \rfloor + \log_2 \left(M\right).
	\label{eq:informationbits3}
\end{equation}
Without loss of generality, $a_{n,m}$ and $\theta_{n,m}$ values are drawn from $M-$QAM constellation points in this paper.

\subsubsection{AN generation}
AN is intentionally generated to safeguard against information leakage across the communication channel connecting WPTx and a potential eavesdropper (Eve) and characterized by the channel matrix $\mathbf{G}_{\text{Eve}}\!\!\!\!\in\!\!\!\!\mathbb{C}^{N_{\text{Eve}} \times N_T \times N}$. The primary aim of introducing this deliberately produced interference is to enhance the robustness of the PLS within the system. Later, we will provide comprehensive insights into how this interference also plays a pivotal role in advancing the energy harvesting process.

{For the AN generation}, it is assumed that  the IR has $N_{\rm \text{IR}}$ receive antennas such that $N_{\rm IR} < N_T$.  Then, the singular value decomposition (SVD) can be implemented through each subband such that $	\textbf{H}_{\text{IR},n}~=~\textbf{U}_n\mathbf{\Lambda}_n\textbf{V}_n^H$  where $\textbf{H}_{\text{IR},n}$ is a $N_{\text{IR}} \times N_T$ channel matrix  with $r={\rm rank}\left(\textbf{H}_{ \text{IR},n}\right)$ and $\textbf{V}_n=\left[\textbf{v}^n_1 \textbf{v}^n_2\dots \textbf{v}^n_r \textbf{v}^n_{r+1} \dots \textbf{v}^n_{N_T}\right]$ includes the nullspaces of $\textbf{H}_{\text{IR},n}$, which are $\textbf{V}_{\perp,n}=\left[\textbf{v}^n_{r+1}  \dots \textbf{v}^n_{N_T}\right]$. Then,  the AN  waveform can  be expressed as 
\begin{equation}
	\pmb{\varepsilon}_n =\sum_{i=r+1}^{N_t}\delta_i \textbf{v}^n_i u_i.
	\label{eq: ANtransmit}
\end{equation}
Herein, \eqref{eq: ANtransmit} is the jamming signal obtained from independent identically distributed (i.i.d.) Gaussian distribution, $u_i\sim\mathcal{CN}\left ( 0,\lambda_u^2 \right )$ along with $\textstyle \sum_{i=r+1}^{N_t}$$\delta_i^2=1$. Since $\textbf{H}_{\text{IR},n} \cdot \textbf{V}_{\perp,n}=0$ holds, the generated AN on top of the existing WPT waveform does not affect on the received signal in IR while it leads to additional jamming power in Eve side due to $\textbf{G}_{\text{Eve},n} \cdot \textbf{V}_{\perp,n}~\ne~0$. Given that AN is generated based on the link between WPTx and IR, it also contributes to additional jamming power on the EH side; hence, $\textbf{G}_{\text{EH},n} \cdot \textbf{V}_{\perp,n} \neq 0$.

As a result, the AN-added received passband waveform at the EH can be expressed as
\begin{equation}
    \mathbf{y}_{\rm EH} = \Re\left\{\sum_{n=1}^{N} \mathbf{G}_{\text{EH},n}\bigg[ \mathbf{w}_n e^{j 2 \pi f_n t} + \pmb{\varepsilon}_n  e^{j 2 \pi f_1 t}\bigg]\right\}.
    \label{eq: R_Signal_EH}
\end{equation}
Herein, the AN waveform aligns with the frequency of the first subband, $f_1$, and its power defined as $\lambda_u^2\slash N$ for $n$th subband  (i.e., $\mathbb{E}[\pmb{\varepsilon}_n^H\pmb{\varepsilon}_n] = \lambda_u^2\slash N$). In \eqref{eq: R_Signal_EH}, AWGN term is omitted, assuming that the antenna noise is negligible and not substantial enough to be harvested.

\subsection{IR: Information harvesting}

The primary objective of the IH mechanism is to extract information from the received WPT-IM waveform. To conduct this, the IR leverages both the mapping rule employed by the IM scheme and the CSI of the WPT-IR link. While energy can be directly harvested from the received RF signal, the IR must acquire the baseband version of the received signal for a specific subband $n$. Following downconversion and filtering processes, the resulting baseband signal at the IR is mathematically expressed as described in \cite{Poor2019}
\begin{equation}
	\mathbf{y}_{\text{IR},n} = \mathbf{H}_{\text{IR},n}\mathbf{x}_n + \mathbf{z}_{\text{IR},n}.
	\label{eq: RTotal_Signal_EH}
\end{equation}
Here, $\mathbf{x}_n$ includes the WPT-IM waveform $\mathbf{w}_n$ along with the AN waveform $\pmb{\varepsilon}_n$ such that $\mathbf{x}_n=\mathbf{w}_n + \pmb{\varepsilon}_n $. In \eqref{eq: RTotal_Signal_EH} $\mathbf{z}_{\text{IR},n}$ denotes the noise vector, where its each element ${z}_{\text{IR},n} \sim \mathcal{CN}\left(0, \sigma^2_n\right)$ has a zero mean and a variance $\sigma^2_n$ for the $n$th subband, and $\mathbf{z}_{\text{IR},n} \!\in \!\mathbb{C}^{N_{\text{IR}} \times 1}$. Consequently, the total received baseband signal at the IR can be represented as:
\begin{equation}
	\mathbf{y}_{\text{IR}} = \sum_{n=1}^{N} \left[\mathbf{H}_{\text{IR},n}\mathbf{x}_n + \mathbf{z}_{\text{IR},n} \right].
\end{equation}

The IR employs a maximum likelihood (ML) detector to estimate transmitted bits from the WPT-IM waveform. Since various IM techniques can be employed for conveying information, the resulting ML representations vary based on the utilized IM scheme. Consequently, the optimum ML decoder representations are provided separately below.

\subsubsection{GSSK/SSK-based IH}
The GSSK technique involves selecting specific columns from each subband component of the channel matrix,  $\mathbf{H}_{\text{IR},n}$ to transmit information bits. Accordingly, the product $\mathbf{H}_{\text{IR},n}\mathbf{x}_n$ yields the term $\mathbf{h}_{\text{IR},n}^l$, which corresponds to the channel coefficient set of the $l$th GSSK codebook out of $L$ possible combinations in total. It is worth noting that in the case of SSK, only a single column of $\mathbf{H}_{\text{IR},n}$ is chosen. Therefore, for mechanisms based on GSSK/SSK, the ML detector is designed to estimate the transmit antenna indices when given the value of $N_a$. The ML procedure can be implemented as follows:
\begin{equation}
	\hat{l}_{\rm IR}=\argmin_{l \in \{1,\dots,L\}} \sum_{n=1}^N\left \| \mathbf{y}_{\text{IR}, n}- \textbf{h}_{\text{IR},n}^l \right \|^2.
	\label{eq:IRdecodingGSSK}
\end{equation}
Here, $\hat{l}_{\rm IR}$ represents the estimated antenna index. 
\subsubsection{GSM/SM-based IH} GSM/SM schemes differ from GSSK/SSK ones through the use of signal modulation, specifically QAM, to enhance spectral efficiency. In cases where the WPT waveform includes a modulated symbol, \eqref{eq:IRdecodingGSSK} can be rewritten as follows:
\begin{equation}
	(\hat{l}_{\rm IR}, \hat{{\omega}})=\argmin _{
		\small l \in \{1,\dots,L\}, \\ \pmb{\omega} \in \psi
	}\sum_{n=1}^{N}\left \| \mathbf{y}_{\text{IR,n}} - \textbf{h}_{\text{IR},n}^l {\omega} \right \|^2.
	\label{eq:IRdecodingGSM}
\end{equation}
{Here, $\omega_{n,m}$ representation in \eqref{eq:mthantennatransmit} simplifies to a ${\omega}$ as IM schemes transmit identical symbols from $N_a$ active antennas across all subbands. Thus, $\omega~=~ ae^{j\theta}$ represents a symbol selected from an $M$-QAM constellation, $\psi$, and $\hat{{\omega}}$ is decoded transmitted symbol.}
\subsubsection{QSSK/GQSSK-based IH}
{Similar to the GSSK/SSK-based mechanism, the ML detector only needs to estimate the active antenna indices without performing symbol decoding. The distinction is that QSSK scheme activates two antennas at a time to convey quadrature components, whereas GQSSK might activate more than two antennas. Hence, the ML detector should jointly estimate in-phase and quadrature components, and it is given by }

\begin{equation}
	\begin{aligned}
		\footnotesize
		&\left( \hat{l}_{\rm IR}^{\Re},\hat{l}_{\rm IR}^{\Im}\right)=\argmin_{\substack{\textbf{h}_{{\rm IR}}^{\Re,l},\textbf{h}_{{\rm IR}}^{\Im,l}  \\l \in \{1,\dots,L\}}}  \sum_{n=1}^N\left \| \mathbf{y}_{\text{IR},n}-\big[\textbf{h}_{\text{IR},n}^{\Re,l} +j\textbf{h}_{\text{IR},n}^{\Im,l} \big]\right \|^2.
	\end{aligned}
	\label{eq:IRdecodingQSSK}
\end{equation}
Herein, ${\mathbf{h}}_{{\rm IR}}^{\Re,l}$ and ${\mathbf{h}}_{\rm IR}^{\Im,l}$  correspond to the $l$th potential codebook channel coefficient set of the WPTx-IR link out of  $L$ candidates for real part and imaginary part transmission, respectively. $\left( \hat{l}_{\rm IR}^{\Re},\hat{l}_{\rm IR}^{\Im}\right)$ refers the estimated indices of active transmit antennas at the IR. For a toy example, $\left( \hat{l}_{\rm IR}^{\Re},\hat{l}_{\rm IR}^{\Im}\right)=\left( \left[1 1 0\right],\left[0 1 1\right]\right)$ implies the $1$st and $2$nd transmit antennas are active for the real part and $2$nd and $3$rd for the imaginary part, respectively.

\subsubsection{GQSM/QSM-based IH}
GQSM/QSM augments QSSK/GQSSK techniques by incorporating $M$-QAM constellation symbols to transmit quadrature components, separately containing the in-phase and quadrature part of the $M$-QAM symbol. If the WPT-IM waveform contains a modulated symbol, \eqref{eq:IRdecodingQSSK} can be rewritten as follows:

\begin{equation}
	\begin{aligned}
		\footnotesize
		&\left(\hat{l}_{\rm IR}^{\Re},\hat{l}_{\rm IR}^{\Im}, \hat{{\omega}}\right)=\\
		&\argmin_{\substack{\textbf{h}_{\rm IR}^{\Re,l},\textbf{h}_{\text{IR},n}^{\Im,l}\\ l \in \{1,\dots,L\}, {\omega}\in \psi}} \sum_{n=1}^N\left \| \mathbf{y}_{\text{IR},n} - \big[\textbf{h}_{\text{IR},n}^{\Re,l} \Re\{{{\omega}}\}+j\textbf{h}_{\text{IR},n}^{\Im,l} \Im\{{{\omega}}\}\big]\right \|^2 ,
	\end{aligned}
	\label{eq:IRdecodingGQSM}
\end{equation}
where $\hat{{\omega}}$ is a decoded transmitted symbol whose real and imaginary parts are carried by different active transmit antennas.

\section{Error Performance Analysis}
In this section, average bit error rates (ABER) at the IR and the Eve are calculated, respectively.
 \subsection{ABER of Information receiver}
The conditional pairwise error probability (CPEP) of deciding on a baseband IM codebook $\pmb{x}_k$ in the case of that  $\pmb{x}_j$ transmitted under ML detection can be calculated by

	\begin{equation}
 \footnotesize
                P({\pmb{x}_j} \rightarrow {\pmb{x}_k} \mid \mathbf{H}_{\rm IR})= 
			    \mathcal{Q}\Bigg(\sqrt{\sum_{i = 1}^{N_{\rm IR}} \frac{\sum_{n = 1}^{N}\lVert \Phi^{(n)}_{\textbf{IR},i}\rVert^2}{2N\sigma _n^2}} \Bigg)  =   \mathcal{Q}\big(\sqrt{\gamma_{\text{IR}}}\big).     \label{eq:CPEP_IH}
	\end{equation}
Herein, $ \Phi^{(n)}_{\textbf{IR},i}$ is defined as $\Phi^{(n)}_{\textbf{IR},i} = (\mathbf{h}^{\Re,j}_{\rm IR,n,i} {\omega}_j^{\Re} - \mathbf{h}^{\Re,k}_{\rm IR,n,i} {\omega}^{\Re}_k) + ( \mathbf{h}^{\Im,j}_{\rm IR,n,i} {\omega}^{\Im}_j - \mathbf{h}^{\Im,k}_{\rm IR,n,i} {\omega}^{\Im}_k) $. Note that \eqref{eq:CPEP_IH} introduces a generalized CPEP expression, and it can be utilized for calculating the different types of the IM schemes, ranging from the ones that do not employ modulated symbols to the ones that also carry a set of modulated symbols. 

\subsubsection{IM schemes with one active Tx per index} To begin with, $\Phi^{(n)}_{\textbf{IR}}$ can be simplified for the SSK and GSSK schemes by considering ${\omega}_j^{\Re} = {\omega}_k^{\Re} = {\omega}_j^{\Im} = {\omega}_k^{\Im} = 1$, and for the QSSK and GQSSK schemes by considering ${\omega}_j^{\Re} = {\omega}_k^{\Re} = 1$ and ${\omega}_j^{\Im} = {\omega}_k^{\Im} = j$. This simplification is applicable to the IM schemes that do not employ signal modulation, namely SSK, GSSK, QSSK, and GQSSK. Consequently, it results in the following expressions: $ \Phi^{(n)}_{\textbf{IR},i} = (\mathbf{h}^{\Re,j}_{\rm IR,n,i} - \mathbf{h}^{\Re,k}_{\rm IR,n,i}) + ( \mathbf{h}^{\Im,j}_{\rm IR,n,i} - \mathbf{h}^{\Im,k}_{\rm IR,n,i}) $ for SSK, and $ \Phi^{(n)}_{\textbf{IR},i} = (\mathbf{h}^{\Re,j}_{\rm IR,n,i} - \mathbf{h}^{\Re,k}_{\rm IR,n,i}) + j( \mathbf{h}^{\Im,j}_{\rm IR,n,i} - \mathbf{h}^{\Im,k}_{\rm IR,n,i}) $ for QSSK. These expressions can subsequently be employed to calculate the ABER performance of both the SSK and QSSK schemes. 

\subsubsection{IM schemes with one multiple Tx per index} In the case of generalized IM schemes that simultaneously activate multiple antennas instead of single one, the corresponding effective channel columns $\mathbf{h}_{\rm IR,n,i}^{j_{\text{eff}}}$ and $\mathbf{h}_{\rm IR,n,i}^{k_{\text{eff}}}$ are considered in the ABER calculation since it represents the summation of the columns corresponding to active transmit antennas in the total channel matrix of  $\mathbf{H}_{\rm IR}$. For instance,  the expression becomes $ \Phi^{(n)}_{\textbf{IR},i} = (\mathbf{h}^{\Re,j_{\text{eff}}}_{\rm IR,n,i} - \mathbf{h}^{\Re,k_{\text{eff}}}_{\rm IR,i }) + ( \mathbf{h}^{\Im,j_{\text{eff}}}_{\rm IR,i } - \mathbf{h}^{\Im,k_{\text{eff}}}_{\rm IR,i }) $  for the GSSK and GQSSK cases. 

\subsubsection{IM schemes with modulated symbol} If there is signal modulation involved, the expression takes the form of $ \Phi^{(n)}_{\textbf{IR}} = (\mathbf{h}^{\Re,j_{\text{eff}}}_{\rm IR,n,i}{\omega}_j^{\Re} - \mathbf{h}^{\Re,k_{\text{eff}}}_{\rm IR,i }{\omega}_k^{\Re}) + ( \mathbf{h}^{\Im,j_{\text{eff}}}_{\rm IR,i } {\omega}_j^{\Im} - \mathbf{h}^{\Im,k_{\text{eff}}}_{\rm IR,i }{\omega}_k^{\Im}) $ for the GSM and GQSM schemes.

Then, \eqref{eq:CPEP_IH} implies that $\gamma_{\text{IR}}$ follows a central chi-squared distribution with $2N_r$ degrees of freedom and parameter $\nu^{\text{IR}}_r$ and it can be written as $ \gamma_{\text{IM}} = \Sigma_{r = 1}^{N_{\rm IR}} |\nu^{\text{IR}}_r|^2$. Herein, $\nu_r =  \Sigma_{n = 1}^{N} w_{n,r}\phi_r\slash{(2N\sigma _n^2)}$ represents the $r$th element of  $\Phi^{(n)}_{\textbf{IR}}$, which has a row length of $N_{\rm IR}$. At this point, the average PEP, $P({\pmb{x}_j} \rightarrow {\pmb{x}_k})$, is written as follows\cite{DiRenzo2010}:

\begin{equation}
	 P({\pmb{x}_j} \rightarrow {\pmb{x}_k}) = \xi^{N_{\rm IR}} \sum_{i=0}^{N_{\rm IR}-1}\left(\begin{array}{c}N_{\rm IR}-1+i \\ i\end{array}\right)\left[1-\xi\right]^i
    \label{eq:PEP_IH}
\end{equation}
where $\xi = (1-\sqrt{(\bar{\nu}\slash 2) \slash(1+\bar{\nu}\slash2) })/2$ and $\bar{\nu}$ represents the mean value of $\nu^{\text{IR}}_r$ as detailed in \cite{Mesleh2015}.

After utilizing \eqref{eq:PEP_IH} and the well-known union bound expression, the generalized ABER expression of the WPT-IM schemes can be calculated from
\begin{equation} 
	\mathrm{ABER} =\frac{1}{\eta} \frac{1}{L} \sum_{j=1}^{L} \sum_{k \neq j=1}^{L} \epsilon\left(j, k\right) P({\pmb{x}_j} \rightarrow {\pmb{x}_k})
    \label{eq:ABER}
\end{equation}
where $ \epsilon(j, k)$ is the number of erroneous bits  when a codeword $\pmb{x}_k$ is decoded at the IR when $\pmb{x}_j$ is transmitted at the WPTx.

 \subsection{ABER of Eavesdropper}

In this subsection, the ABER expression at the Eve is investigated and the results herein can be utilized for evaluating a potential case where the EH acts as a malicious role such that EH tries to intercept information transmission intended for an information receiver for a given service area. 

To begin with, as similar to \eqref{eq:CPEP_IH}, the conditional PEP expression for Eve can be expressed as follows:

	\begin{equation}
 \footnotesize
                P({\pmb{x}_j} \rightarrow {\pmb{x}_k} \mid \mathbf{G}_{\text{Eve}})= 
			    \mathcal{Q}\Bigg(\sqrt{\sum_{i = 1}^{N_{\rm Eve}} \frac{\sum_{n = 1}^{N}\lVert \Psi_n\hat{\Phi}^{(n)}_{\textbf{Eve},i}\rVert^2}{2N\sigma _n^2}} \Bigg)  =   \mathcal{Q}\big(\sqrt{\gamma_{\text{Eve}}}\big).        \label{eq:CPEP_EH}
	\end{equation}
Herein, $\hat{\Phi}^{(n)}_{\textbf{Eve},i}$ is defined as $\hat{\Phi}^{(n)}_{\textbf{Eve},i} = (\mathbf{g}^{\Re,j}_{\rm Eve,n,i} \omega_j^{\Re} - \mathbf{g}^{\Re,k}_{\rm Eve,n,i} \omega^{\Re}_k) + ( \mathbf{g}^{\Im,j}_{\rm Eve,n,i} \omega^{\Im}_j - \mathbf{g}^{\Im,k}_{\rm Eve,n,i} \omega^{\Im}_k) $   and $\Psi_n$ = $\sigma_n\mathbf{C}_{\epsilon}^{-1\slash2}$ stands for the whitening transformation matrix. Note that the power of $\Psi_n$ needs to be arranged based on the total number of available subbands and after multiplying $\Psi_n$ with given WPT-IM waveforms, \eqref{eq:CPEP_EH} reduces to

\begin{equation}
    \gamma_{\text{Eve}} =  {\sum_{i = 1}^{N_{\rm Eve}} \frac{\sum_{n = 1}^{N}\lVert \mathbf{w}_n\Phi^{(n)}_{\text{Eve}}\rVert^2}{2(N\sigma _n^2 +\lambda_u^2)}}.
\end{equation}
Herein, $\gamma_{\text{Eve}} = \Sigma_{r = 1}^{N_{\rm Eve}} |\nu^{\text{Eve}}_r|^2$ shows the similar characteristics with $\gamma_{\text{IR}}$ given in \eqref{eq:CPEP_IH} and the parameter $\nu^{\text{Eve}}_r = \Sigma_{n = 1}^{N} w_{n,r}\phi_r\slash{(2N\sigma _n^2 + 2\lambda_u^2)}$ signifies the $r$-th element of the vector $\Phi^{(n)}_{\textbf{Eve}}$, which has a  row length of $N_{\rm IR}$. In this regard, the selection of $\Phi^{(n)}_{\textbf{Eve}}$ should be aligned with the selected IM schemes for information harvesting, as in the selection of $\Phi^{(n)}_{\textbf{IR}}$.
Then, the calculation of the ABER at the Eve can be obtained after substituting \eqref{eq:CPEP_EH} into \eqref{eq:ABER}.

\section{Ergodic Secrecy Rate Analysis}

In this section, we provide a generalized analytical framework of ergodic secrecy rate for the proposed IM-based information harvesting schemes.
To do so, GQSSK-based IH mechanism was considered. In this respect, each transmit antenna set for the real and the imaginary parts is selected with the same probability, $\frac{1}{L}$. Then, the received signal at the IR through $N$ band, $\mathbf{y}_{\rm IR}$,  obeys the following distribution  \cite{Huang2022}
\begin{equation}
	p \left(\mathbf{y}_{\rm IR}\right)=\frac{1}{L^2}\sum_{j=1}^{L}\sum_{k=1}^{L} \frac{1}{\pi \sigma_n^2} e^{-\frac{\lVert\mathbf{r}_{\rm IR}\rVert^2}{\sigma_n^2}}.
	\label{eq:ProbBob}
\end{equation}
Here, $\mathbf{r}_{\rm IR}=\mathbf{y}_{\rm IR}- \textbf{h}_{\rm IR}^{\Re,j} \Re\{{\omega}\}-j\textbf{h}_{\rm IR}^{\Im,k} \Im\{{\omega}\}$ for a transmitted WPT signal, $\omega$. Then, the mutual information, $\mathcal{I}_{\rm IR}\left(\textbf{r}_{\rm IR} \right)$,  can be expressed in \eqref{eq:MutualInfoIR}, which is a special case of [Eq. (14), \cite{Huang2022}]. Therein, $\textbf{d}_{l_1, l_2}^{r_1, r_2}$  is defined as:
\begin{figure*}
	\begin{equation}
		{
			\begin{aligned}
				&\mathcal{I}_{\rm IR}\left(\textbf{y}_{\rm IR}\right)=\log_2\left(L^2\right)-\frac{1}{L^2}\sum_{l_1=1}^{L}\sum_{l_2=1}^{L}\mathop{\mathbb{E}}\hspace{0.05em}_{\mathbf{z}_{\text{IR}}}\left[\log_2\left(\sum_{r_1=1}^{L}\sum_{r_2=1}^{L}e^{-\frac{\begin{Vmatrix}
							\textbf{d}_{l_1, l_2}^{r_1, r_2}+\textbf{z}_{\rm IR}\end{Vmatrix}^2-\,\begin{Vmatrix}
							\textbf{z}_{\rm IR}\end{Vmatrix}^2}{\sigma_n^2}}\right)\right]. 
			\end{aligned}
		}
		\label{eq:MutualInfoIR}
	\end{equation}
\end{figure*}
\begin{figure*}
	\begin{equation}
		{
			\begin{aligned}
				&\mathcal{I}_{\rm Eve}\left(\textbf{y}_{\rm Eve}\right)=\log_2\left(L^2\right)-\frac{1}{L^2}\sum_{l_1=1}^{L}\sum_{l_2=1}^{L}\mathop{\mathbb{E}} \hspace{0.05em}_{\mathbf{z}_{\text{Eve}}}\left[\log_2\left(\sum_{r_1=1}^{L}\sum_{r_2=1}^{L}e^{-\frac{\begin{Vmatrix}
							\pmb{\delta}_{l_1, l_2}^{r_1, r_2}+\textbf{z}_{\rm Eve}\end{Vmatrix}^2-\,\begin{Vmatrix}
							\textbf{z}_{\rm Eve}\end{Vmatrix}^2}{\sigma_n^2}}\right)\right]. 
		\end{aligned}}
		\label{eq:MutualInfoEve}
	\end{equation}  \vspace{-2em}
	
\end{figure*}

\begin{equation}
	\footnotesize
	\textbf{d}_{l_1, l_2}^{r_1, r_2}=\textbf{h}_{\rm IR}^{\Re,{l_2}_{\rm eff}}\Re\{\omega\}+j\textbf{h}_{\rm IR}^{\Im,{l_1}_{\rm eff}}\Im\{\omega\}-\textbf{h}_{\rm IR}^{\Re,{r_1}_{\rm eff}}\Re\{\omega\}-j\textbf{h}_{\rm IR}^{\Im,{r_2}_{\rm eff}}\Im\{\omega\}
	\label{eq:dl1dl2}
\end{equation}
where  $\textbf{h}_{\rm IR}^{\rm eff} $ is the effective channel after incorporating only active transmit antenna set at the IR such that $\textbf{h}_{\text{Eve}}^{i_{\rm eff}}=\textbf{h}_{\text{Eve}}^{i,1}+ \dots +\textbf{h}_{\text{Eve}}^{i,N_a} $ \cite{yaman2015}. Similar to \eqref{eq:ProbBob}, the received signal at the Eve obeys the following distribution  \cite{Huang2022}
\begin{equation}
	p \left(\mathbf{y}_{\rm Eve}\right)=\frac{1}{L^2}\sum_{j=1}^{L}\sum_{k=1}^{L}\frac{1}{\pi \sigma_n^2} e^{-\frac{\lVert\mathbf{r}_{\rm Eve}\rVert^2}{\sigma_n^2}}.
	\label{eq:ProbEve}
\end{equation}
Therein, $\mathbf{r}_{\rm Eve}=\mathbf{y}_{\rm IR}- \textbf{g}_{\rm Eve}^{\Re,j} \Re\{{\omega}\}-j\textbf{g}_{\rm Eve}^{\Im,k} \Im\{{\omega}\}$. After accounting for the existence of the AN at the Eve side, the mutual information at Eve, $\mathcal{I}_{\text{Eve}}\left(\mathbf{r}_{\text{Eve}} \right)$, is given in \eqref{eq:MutualInfoEve}, which is a special case of [Eq. (15), \cite{Huang2022}]. Here, $\mathbf{z}_{\text{Eve}}$ is the Gaussian noise at the Eve and $\pmb{\delta}_{l_1, l_2}^{r_1, r_2}$ can be expressed as:
\begin{equation}
	\begin{aligned}
		\pmb{\delta}_{l_1, l_2}^{r_1, r_2}=&	\mathbf{C}_{\epsilon}^{-\frac{1}{2}}\left(\textbf{g}_{\text{Eve}}^{\Re,{l_2}_{\rm eff}}\Re\{\omega\}+j\textbf{g}_{\text{Eve}}^{\Im,{l_1}_{\rm eff}}\Im\{\omega\}\right)\\
		&-\mathbf{C}_{\epsilon}^{-\frac{1}{2}}\left(\textbf{g}_{\text{Eve}}^{\Re,{r_1}_{\rm eff}}\Re\{\omega\}-j\textbf{g}_{\text{Eve}}^{\Im,{r_2}_{\rm eff}}\Im\{\omega\}\right),
	\end{aligned}
\end{equation}
where $\textbf{g}_{\text{Eve}}^{i_{\rm eff}}$ is the effective channel and $	\mathbf{C}_{\epsilon}$ refers a covariance matrix of interference plus noise term at Eve, which is,
\begin{equation}
	\mathbf{C}_{\epsilon}=\frac{\lambda_u^2}{N(N_T-r)}\mathbf{G}_{\rm Eve,n}\left(\sum_{i=r+1}^{N_T}\textbf{v}_i^{(n)} \textbf{v}_i^{(n)^H}\right) \mathbf{G}_{\rm Eve,n}^H+\sigma_n^2\mathbf{I}.
\end{equation}
Note that the channel between the WPTx and the IR relies on GQSSK-based modulated antenna indices so the capacity analysis deducts into a capacity analysis for the discrete-input continuous-output memoryless channel (DCMC) \cite{singh2021} and it might be not a straightforward to obtain the closed-form analysis in most cases. In this respect, the secrecy rate of IR can be expressed as \cite{yaman2015}
\begin{equation}
	\mathcal{R}_{\rm IR}=\max \{0, \,\, \mathcal{I}_{\rm IR}\left(\textbf{y}_{\rm IR} \right)-\mathcal{I}_{\rm Eve}\left(\textbf{y}_{\rm Eve} \right)\},
	\label{eq:SecrecyRate}
\end{equation}
Note that positive secrecy from \eqref{eq:SecrecyRate} implies communication opportunity on top of the existing WPT mechanism, even if some information can be leaked to Eve in the service area.

For the other IM schemes given in Section II.B, the ergodic secrecy rates can be obtained from \eqref{eq:SecrecyRate} after modifying \eqref{eq:MutualInfoIR} and \eqref{eq:MutualInfoEve}. For instance, in the case of GQSM/QSM-based IH, the summation limits should be replaced with $L \to L M$  where $M$ is the modulation order and $\omega$ seen in  \eqref{eq:dl1dl2} becomes a variable of the summation indices rather than constant waveform such that $\omega_{l_1}$. 
For GQSSK/SSK schemes, the secrecy calculation requires one effective channel definition without dividing real and imaginary parts so the definition of $\mathbf{r}_{\rm IR}$ in \eqref{eq:ProbBob} is reformulated as $\mathbf{r}_{\rm IR,n}=\mathbf{y}_{\rm IR,n}-\textbf{h}^n_{j} \omega$ \cite{yaman2015} which results in 

\begin{equation}
	\begin{aligned}
		&\mathcal{I}_{\rm IR}\left(\textbf{y}_{\rm IR}\right) = \log_2\left(L\right)\\
		&-\frac{1}{L}\sum_{l_1=1}^{L}\mathop{\mathbb{E}} \hspace{0.05em}_{\textbf{z}_{\rm IR}}\left[\log_2\left(\sum_{l_2=1}^{L}e^{-\frac{\begin{Vmatrix}
					\textbf{d}_{l_1}^{l_2}+\textbf{z}_{\rm IR}\end{Vmatrix}^2-\,\begin{Vmatrix}
					\textbf{z}_{\rm IR,n}\end{Vmatrix}^2}{\sigma_n^2}}\right)\right]. 
	\end{aligned}
\end{equation}

A similar approach is functional when considering the scenario involving Eve, where its mutual information can be expressed as follows:

\begin{equation}
	\begin{aligned}
		&\mathcal{I}_{\rm Eve}\left(\textbf{y}_{\rm Eve} \right) = \log_2\left(L\right)\\
		&-\frac{1}{L}\sum_{l_1=1}^{L}\mathop{\mathbb{E}} \hspace{0.05em}_{\textbf{z}_{\rm Eve}}\left[\log_2\left(\sum_{l_2=1}^{L}e^{-\frac{\begin{Vmatrix}
					{\pmb{\varrho}}_{l_1}^{l_2}+\textbf{z}_{\rm Eve}\end{Vmatrix}^2-\,\begin{Vmatrix} \textbf{z}_{\rm Eve}\end{Vmatrix}^2}{\sigma_n^2}} \right)\right].
	\end{aligned}
\end{equation}
Herein, $\textbf{d}_{l_1}^{l_2}$ and $\pmb{\varrho}_{l_1}^{l_2}$ correspond to the channel distances between $m_1$th and $m_2$th antenna index combinations at the IR and the Eve for the $n$th subband, where $N_a$ antennas out of $N_T$ are active such that  $\textbf{d}_{l_1}^{l_2}~=~\textbf{h}_{\rm l_1, eff}-\textbf{h}_{\rm l_2, eff}$ and $\pmb{\varrho}_{l_1}^{l_2}=\textbf{g}_{\rm l_1, eff}-\textbf{g}_{\rm l_2, eff}$, respectively.   For GSM/SM-based IH, the summation limits should be replaced with $L \to L M$  where $M$ is the modulation order,  the calculation of  $\textbf{d}_{l_1}^{l_2}$ takes into account $\omega$ as in \cite{yaman2015}.
	
\section{Numerical Results}

\begin{table}[b!]
\captionsetup{justification=centering, labelsep=newline}
\caption{SIMULATION PARAMETERS USED IN IM-BASED IH MECHANISM.}
\footnotesize
\begin{center}    
\resizebox{0.9\columnwidth}{!}{%
\begin{tabular}{|c|c|c|}
\hline
\text{Parameter} & \text{Value}                                                  & \text{Reference}              \\ \hline
$\{k_2,k_4\}$      & \begin{tabular}[c]{@{}c@{}}$\{0.0034, 0.3829\}$\end{tabular} & \cite{Kim2020} \\ \hline
Path-loss model    & $35.3 + 37.6 \log_{10}(d)$                                      & \cite{Ilter2022GL}  \\ \hline
$R_{\text{ant}}$   & $50$ $\Omega$                                                   & \cite{Kim2020} \\ \hline
$P_T$              & $36$ dBm                                                        & \cite{FCC}    \\ \hline
$\rho$             & $\left[0, 1\right]$                                             & -                               \\ \hline
$N$                & $\{1, 3, 5\}$                                                   & -                               \\ \hline
\end{tabular}%
}
\end{center}
\label{tab:WPTparameter}
\end{table}

In this section, we investigate the feasibility of  IM-based IH schemes in terms of their energy harvesting capability, bit error rate (BER) performance, and ergodic secrecy rates (ESR) under different scenarios where the multitone signals are used during WPT transmissions. To do so, the channels between WPTx and EH, as well as between the WPTx and IR, are assumed Rayleigh fading. The simulation parameters regarding WPTx are given in Table~\ref{tab:WPTparameter}. Since the EH's ability to scavenge energy is directly linked to the distance between the transmitter and receiver \cite{Ho2012}, the following path loss model, formulated as  $35.3 + 37.6 \log_{10}(d)$, where $d$ is the distance between WPTx and EH, is considered. In computer simulations, the EH is assumed to consist of $N_{\text{EH}}$ pair of receive antennas and rectennas. In rectennas, it is assumed that have perfect matching and an ideal low-pass filter, and $R_{\text{ant}}$ is set to 50 $\Omega$. The parameters $k_2$ and $k_4$ are selected as $0.0034$ and $0.3829$, respectively. In order to limit the maximum effective isotropic radiated power (EIRP) as specified in FCC Title 47, Part 15 regulations \cite{FCC}, the total transmit power is considered as $P_T = 36$ dBm at the WPTx. 

In a scenario where Eve possesses knowledge of the mapping rule utilized by the IM schemes, it could intercept the information transmission between WPTx and IR. In this situation, Eve is capable of conducting information harvesting, thus enabling it to decode the information originally intended for Bob. Therefore, to counter Eve's interception AN is added to the transmitted WPT-IM waveform. The overall transmit power is divided into two parts, which are AN and IM, based on the power allocation factor, $\rho$, ($0 \leq \rho \leq 1$). Specifically, $\lambda^2_u = \rho P_T$ represents the transmit power of the AN waveform, and $\lambda^2_s = (1-\rho)P_T$ represents the transmit power of the IM waveform. Consequently, as $\rho$ increases, the transmit power allocated for the IM waveform decreases, and that of the AN waveform increases. 
 
 The WPTx emits multitone IM waveforms where the available power is allocated equally between $N_a$ active transmit antennas, whereas the generated AN which is emitted from all antennas. In order to boost the harvested DC power, the IM waveforms are combined with a multitone signal vector $\mathbf{w}_n$. Note that WPTx-IR waveforms do not rely on CSI, although the CSI of the WPTx-IR link is available at the transmitter, and it is utilized in generating the AN. Multi-tone signals with a bandwidth denoted as $B$, which is determined by the expression $(n-1)\Delta{f_N}$ where $\Delta{f_N}$ corresponds to the inter-carrier frequency spacing, equivalent to 1 kHz, are utilized in simulations. The frequency of the first subband, $f_1$, is set at 100 kHz, while the subsequent subbands adhere to the frequency relationship $f_N = f_1 + \Delta f_N$.

 \subsection{Energy harvesting capability}
 
\begin{figure}[!t]
		\centering
		\includegraphics[width=0.46\textwidth]{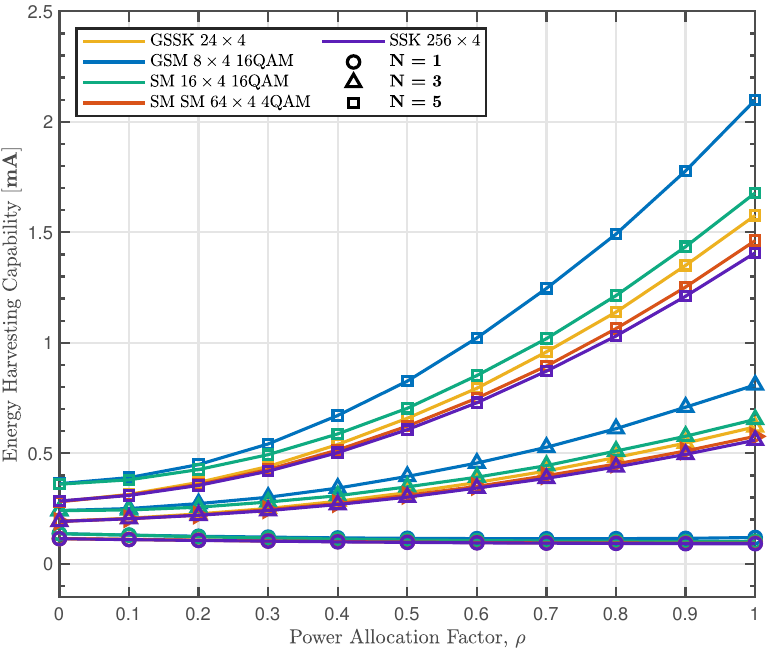}
		\caption{The energy harvesting capability ($z_{DC}$) comparison of SM schemes with $\eta = 8$ for different $N$ values with respect to varying power allocation factors, the EH is located at $d=$ 1.5$m$ from the WPTx.}
		\label{fig:zDC_SM}
\end{figure}

\begin{figure}[!t]
		\centering
		\includegraphics[width=0.46\textwidth]{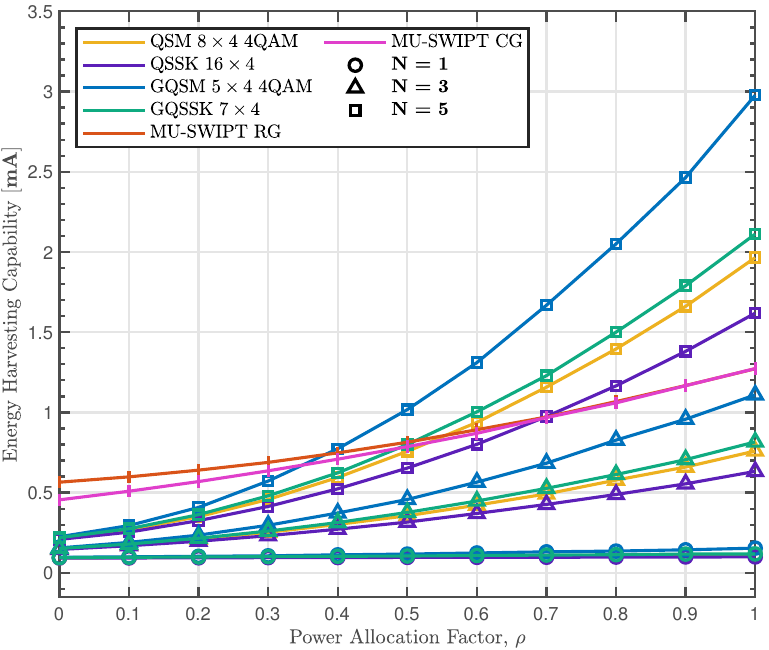}
		\caption{The energy harvesting capability ($z_{DC}$) comparison of  QSM schemes with $\eta = 8$ for different $N$ values with respect to varying power allocation factors, the EH is located at $d=$ 1.5$m$ from the WPTx.}
		\label{fig:zDC_QSM}
\end{figure}

To begin with, we aim to explore the benefits of the IM-based IH mechanism at the EH. To do so, Figs. {\ref{fig:zDC_SM}} and {\ref{fig:zDC_QSM}} present comparative analysis, and they provide an exhaustive evaluation of the energy harvesting capabilities with varying parameters, different values of $N$ and $\rho$. The comparison involves diverse IM schemes, which are GSSK with $N_T=24$, GSM with $N_T=8$ and 16-QAM, SM with $N_T=16$  and 16-QAM, and also SM with $N_T=64$  and 4-QAM. Similarly, \figurename{ \ref{fig:zDC_QSM}} comprises QSM with $N_T=8$  and 4-QAM, QSSK with $N_T$~=~16, GQSM with $N_T=5$ and 4-QAM, and GQSSK with $N_T=7$ and 4-QAM.  Note that all IM schemes share a common spectral efficiency of $\eta = 8$ and in all scenarios, the parameters $N_a$ and $N_{\text{EH}}$ are set to 2 and 4, respectively. Notably, all IM schemes exhibit nearly identical results in the case of CW signals ($N = 1$) for RF energy harvesting.

 In \figurename{ \ref{fig:zDC_SM}}, a comparison is also conducted between SM schemes utilizing 4-QAM and 16-QAM. This comparison is aimed at illustrating how the nonlinearity of the rectifier is influenced by modulations with larger high-order moments. Notably, the modulation type selected has an impact on the system's energy harvesting capability, as also observed in \cite{Cansiz2020}. Specifically, it is observed that 16-QAM outperforms 4-QAM in terms of energy harvesting capability due to higher $M$ values inducing amplitude fluctuations, in turn, substantially enhance the efficiency of RF-to-DC conversion within the rectifier \cite{Bayguzina2016}. In contrast, modulation schemes like 4-QAM/PSK are unable to attain the same level of efficiency due to their characteristic constant envelope behavior, as mentioned in \cite{Clerckx2022}.
 
Another crucial factor significantly affecting the energy harvesting capability $z_{DC}$ is the power allocation factor, $\rho$. The introduction of AN in the IH system results in interference within the EH receiver. The highest attainable value of $z_{DC}$ is achieved when $\rho$ equals 1, which implies the exclusive transmission of AN without any concurrent information-bearing signals. This distinct interference phenomenon brings strategic benefits to EH, as assigning a greater portion of power to AN generation enhances the energy harvesting capability.

Moreover, increasing the number of antennas in the WPTx system does not lead to enhanced energy harvesting capability due to the selected signaling scheme, which does not rely on CSI. For instance, when comparing SSK with QSSK at the same spectral efficiency, QSSK utilizes $16$ antennas, while SSK employs $256$ antennas in the WPTx system. However, QSSK, which activates only two antennas to convey quadrature components, outperforms SSK, which uses only one transmit antenna to convey information at a time. Additionally, from the information harvesting perspective, increasing the number of transmit antennas enhances the spectral efficiency of the IH system, enabling the transmission of more information. Nonetheless, the energy harvesting performance of the IH system remains unaffected by the increase in the number of antennas.

{\subsubsection*{ Comparison with MU-SWIPT}
In addition to the comparisons between different IM techniques,  Fig. \ref{fig:zDC_QSM} illustrates a comparative analysis of IH systems and the MU-SWIPT system \cite{Zhang2014} in terms of energy harvesting capability. In the MU-SWIPT setting,  $N_T = 6$ transmit antennas at the access point (AP) and a single receive antenna at both the IR and energy receiver (ER) are assumed along with matched filtering (MF) \cite{Kim2020} for both the AP-IR and AP-ER links. Specifically, CSCG inputs are assumed for information signals, while complex Gaussian (CG) and real Gaussian (RG) signals are employed for energy signals. The power allocation is evenly distributed between the information and energy signals, along with the incorporation of a jamming signal into the transmit signal vector.} {As depicted in Fig. \ref{fig:zDC_QSM}, the IH system outperforms the considered MU-SWIPT architecture as the parameter $\rho$ increases, particularly in the scenario with $N = 5$. Within the MU-SWIPT system, allocating $P_T$ between information and energy signals is required, both of which are transmitted across all the transmit antennas. Consequently, deploying MF demands an equivalent number of RF chains as the number of transmit antennas, leading to increased power consumption at the transmitter.
Meanwhile, the power consumption of the IH mechanism remains unaffected with respect to the total number of antennas. Therefore, fewer RF chains can be enough to leverage the benefits of IM schemes, contributing to a simpler and power efficient transmitter design \cite{Basar2017}. }

\subsection{Error performance}

\begin{figure}[!t]
	\centering
	\includegraphics[width=0.46\textwidth]{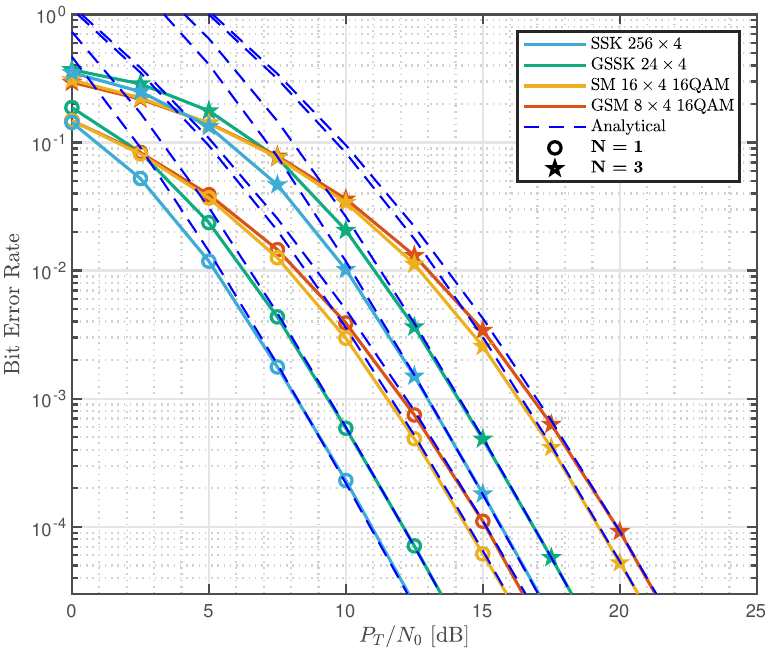}
	\caption{The ABER comparison at the IR for SM schemes with $\eta = 8$, $\rho = 0.2$ and for varying $N$ parameters.}
	\label{fig:BER_SM}
\end{figure}

\begin{figure}[!t]
	\centering
	\includegraphics[width=0.46\textwidth]{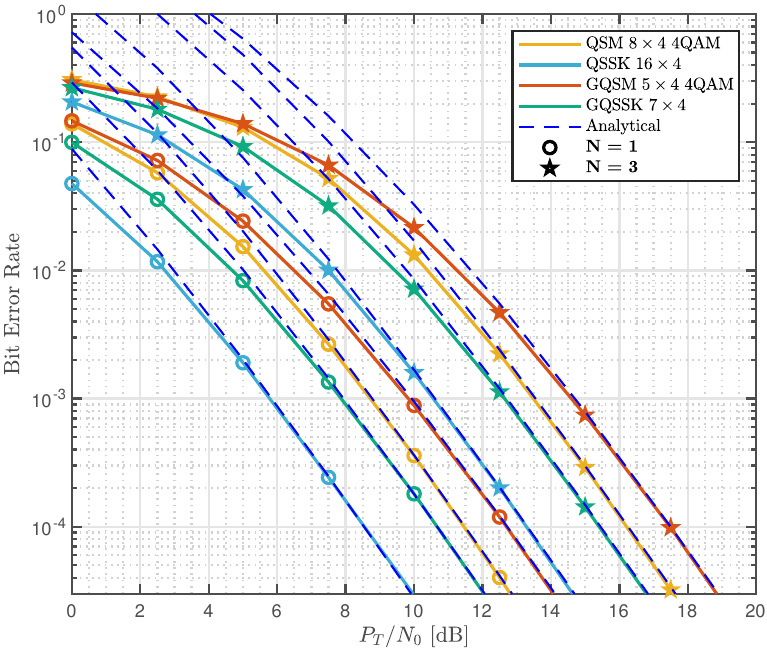}
	\caption{The ABER comparison at the IR for SM schemes with $\eta = 8$, $\rho = 0.2$ and for varying $N$ parameters.}
	\label{fig:BER_QSM}
\end{figure}

In this subsection, we investigate the error performance of the proposed IM-based IH mechanism and provide numerical results to support its merits. In order to demonstrate the impact of multitone signals on the information harvesting performance of IH, the BER of the WPT-IM schemes is plotted in Figs. {\ref{fig:BER_SM}} and \ref{fig:BER_QSM} with the similar IM schemes given in \figurename{ \ref{fig:zDC_SM}} and \figurename{ \ref{fig:zDC_QSM}}, respectively. In WPT-IM schemes, each subband experiences the same channel gains per antenna between the WPTx and IR. Furthermore, noise is added to subbands separately for each receive antenna, as the increase in $N$ results in more noise in the received signal vector. This, in turn, has a detrimental effect during the information decoding process. The simulation results are validated with analytical results given in \eqref{eq:ABER} for both IR and EH. As expected, the analytical curves become tighter as $P_T/N_0$ increases.  The BER of each IM scheme for single subband transmission is approximately 5 dB better than in the case where $N = 3$.  Hence, there is a non-trivial tradeoff between the $z_{DC}$ and BER performance of the information harvesting mechanism. 

Figs. { \ref{fig:BER_SM}} and {\ref{fig:BER_QSM}} also illustrate that best performance results are attained when all bits are modulated in the spatial domain after SSK  and QSSK consistently exhibit superior results compared to the others. Thus, increasing the number of modulated bits in the spatial domain generally enhances the performance of IM schemes. Nonetheless, generalized variants of these techniques, such as GSSK, GQSM, and GQSSK, yield inferior outcomes due to the fact that generalized IM techniques introduce spatial correlation in the spatial domain, adversely affecting the IH system's performance in the end.

\begin{figure}[!t]

\begin{center}
    \begin{subfigure}{0.46\textwidth}
    \includegraphics[width=\linewidth]{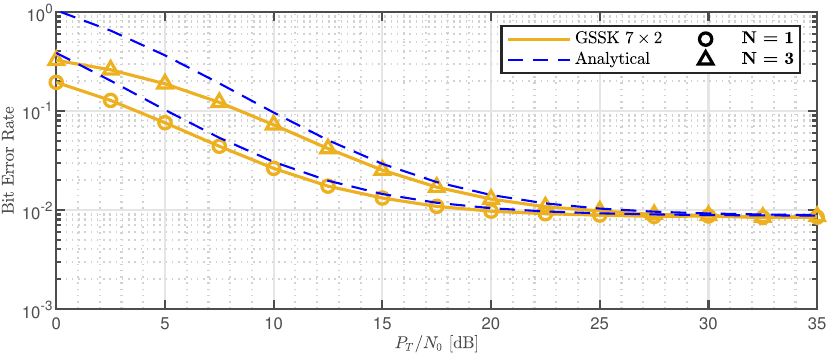}
    \end{subfigure}
\end{center}

\begin{center}
\vspace{-0.2cm}

    \begin{subfigure}{0.46\textwidth}
    \includegraphics[width=\linewidth]{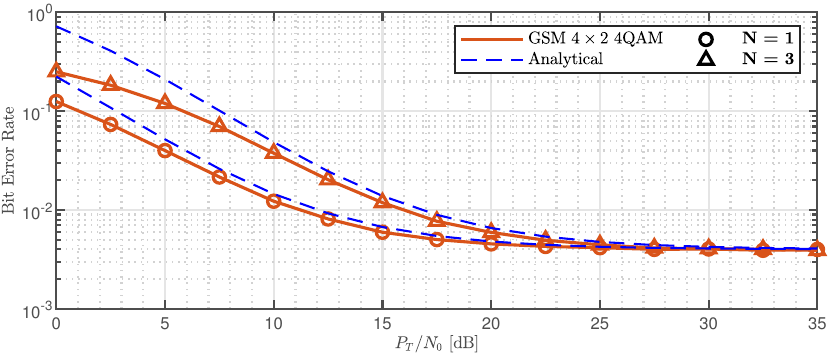}
    \end{subfigure}
\end{center}

\begin{center}
\vspace{-0.2cm}

    \begin{subfigure}{0.46\textwidth}
    \includegraphics[width=\linewidth]{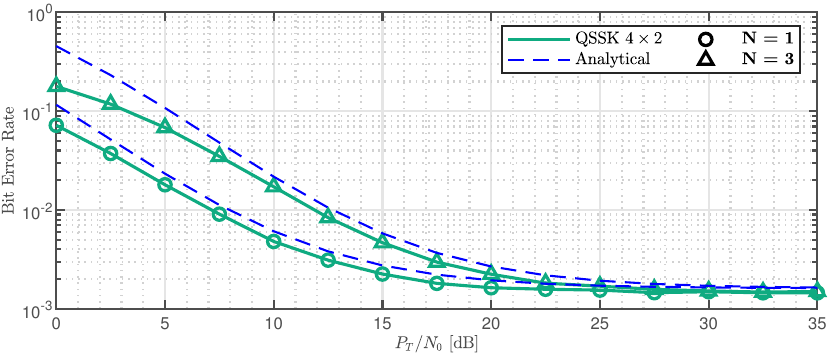}
    \end{subfigure}

\end{center}
\caption{The ABER comparison of Eve with $\eta = 4$, $\rho = 0.025$ and for varying $N$ parameters.}\label{fig:BER_Eve}

\end{figure}

The impact of AN on Eve's BER performance is particularly investigated in \figurename{ \ref{fig:BER_Eve}}, considering the use of GSSK, GSM, and QSSK schemes with $N = 1$ and $N = 3$. The alignment of the analytical curves becomes more apparent as the values of $P_T/N_0$ increase. Specifically, within the high $P_T/N_0$ range, particularly when $P_T/N_0$ exceeds 5 dB, the simulation and analytical curves exhibit a close match. It is observed that the presence of AN disrupts Eve's ability to decode the transmitted data, leading to a substantial deterioration in Eve's BER performance due to the introduced interference. This observation highlights the protective effect of larger subbands against eavesdropping at Eve's end.

\subsection{Secrecy analysis}

\begin{figure}[!b]
	\centering
	\includegraphics[width=0.46\textwidth]{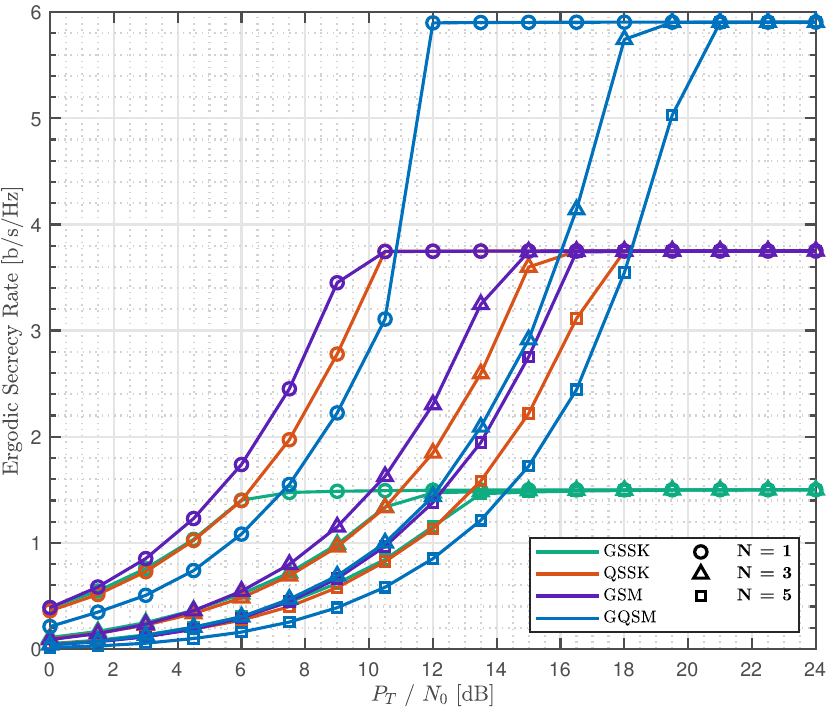}
	\caption{The ESR of WPT-IM schemes \eqref{eq:SecrecyRate} for $\rho$ = 0.2 and for varying $N$ values.}
	\label{fig:ESR03}
\end{figure}

\begin{figure}[!tb]
	\centering
	\includegraphics[width=0.46\textwidth]{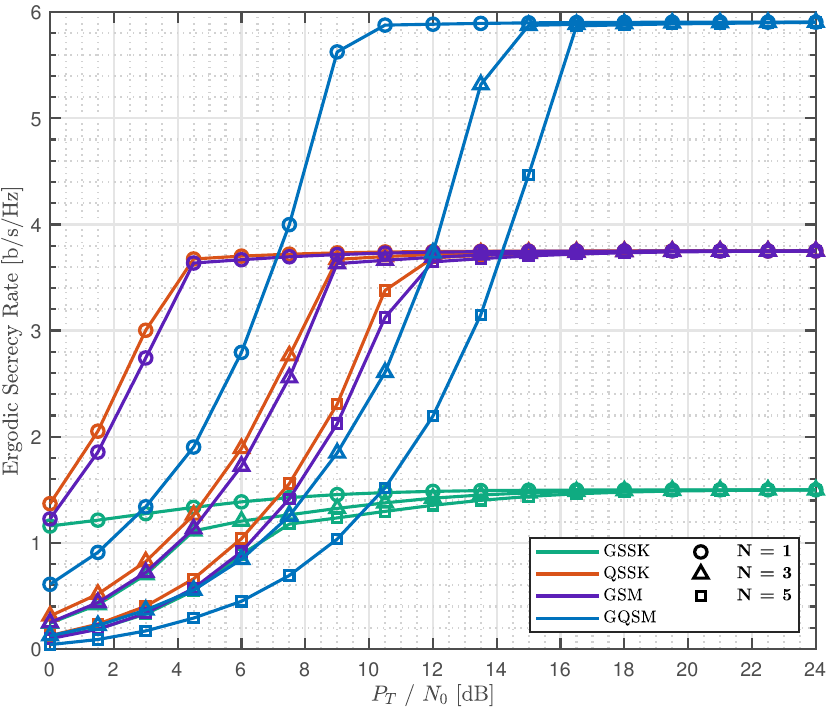}
	\caption{The ESR of WPT-IM schemes \eqref{eq:SecrecyRate} for $\rho$ = 0.6 and for varying $N$ values.}
	\label{fig:ESR06}
\end{figure}

Now, the ESR of the proposed IM-based IH schemes is illustrated with respect to different power allocation scenarios. Figs. {\ref{fig:ESR03}} and {\ref{fig:ESR06}} present the analysis of ESR for GSSK, QSSK, QSM, and GQSM schemes, incorporating multitone signals where  $\rho = 0.2$ and $\rho = 0.6$, respectively. The ESR evaluation for the WPT-IM schemes is conducted under the same conditions, where $N_T = 4$, $N_a = 2$, $N_{\rm IR} = 2$, and 4-QAM is used when necessary, leading to varying spectral efficiencies among the schemes. It is observed that for the case of $\rho = 0.2$, the GSM scheme provides higher secrecy compared to the other ones, while for $\rho = 0.6$, QSSK outperforms all the schemes over low $P_T/N_0$ values with the expense of higher decoding complexity. In the high $P_T/N_0$ region, the GQSM scheme yields the best performance.

Also, allocating more power to generating AN significantly enhances the secrecy performance, so it prevents the eavesdropper from detecting injected data in WPT-IM waveform during information seeding. Additionally, the ESR performance of the WPT-IM schemes is adversely affected by an increasing number of subbands. Consequently, employing a higher number of subbands proves to be advantageous for wireless power transmission. However, it is crucial to consider the cumulative impact of channel gains and of Gaussian noise at the receiver side, both of which can potentially lead to a degradation in system performance in terms of wireless information transmission.

\section{Concluding remarks}
As we transition into the era of 6G technology, addressing the challenge of powering IoT devices with limited or no battery capacity becomes paramount. Energy harvesting technology, particularly far-field WPT, appears to be a promising solution to achieve zero-power communication goals. Consequently, we anticipate a proliferation of combinations of WPT and WIT mechanisms in the near future. These developments are instrumental in realizing green communication architectures that significantly reduce dependence on battery levels and lifetimes. Within this context, a novel protocol known as Information Harvesting (IH) has emerged. IH leverages existing WPT mechanisms to facilitate wireless communication, offering a promising avenue for powering and enabling IoT devices.

This paper has presented a comprehensive framework for the IH mechanism, incorporating various IM techniques. Our simulations and analytical work have demonstrated the advantages of IM-based IH mechanisms, particularly in terms of energy harvesting capabilities at the EH and the reliability of communication at the IR. In summary, IH introduces a novel communication channel atop existing WPT mechanisms, which holds great potential for IoT devices in the 6G era.

\section*{Acknowledgment}
This study has been supported by the Academy of Finland (grant number 334000).
\bibliographystyle{IEEEtran}
\bibliography{Journal2023bib}

\begin{thebibliography}{10}
\providecommand{\url}[1]{#1}
\csname url@samestyle\endcsname
\providecommand{\newblock}{\relax}
\providecommand{\bibinfo}[2]{#2}
\providecommand{\BIBentrySTDinterwordspacing}{\spaceskip=0pt\relax}
\providecommand{\BIBentryALTinterwordstretchfactor}{4}
\providecommand{\BIBentryALTinterwordspacing}{\spaceskip=\fontdimen2\font plus
\BIBentryALTinterwordstretchfactor\fontdimen3\font minus
  \fontdimen4\font\relax}
\providecommand{\BIBforeignlanguage}[2]{{%
\expandafter\ifx\csname l@#1\endcsname\relax
\typeout{** WARNING: IEEEtran.bst: No hyphenation pattern has been}%
\typeout{** loaded for the language `#1'. Using the pattern for}%
\typeout{** the default language instead.}%
\else
\language=\csname l@#1\endcsname
\fi
#2}}
\providecommand{\BIBdecl}{\relax}
\BIBdecl

\bibitem{Zhang2022}
Y.~Zhang and B.~Long, ``A review of {5G}-advanced service and system aspects
  standardization in {3GPP},'' in \emph{2022 IEEE/CIC Int. Conf. Commun. in
  China (ICCC Workshops)}, 2022, pp. 94--99.

\bibitem{Naser2023}
\BIBentryALTinterwordspacing
S.~Naser, L.~Bariah, S.~Muhaidat, and E.~Basar, ``Zero-energy devices empowered
  {6G} networks: Opportunities, key technologies, and challenges,'' 2023.
  [Online]. Available:
  \url{https://www.techrxiv.org/articles/preprint/Zero-Energy_Devices_Empowered_6G_Networks_Opportunities_Key_Technologies_and_Challenges/21558030}
\BIBentrySTDinterwordspacing

\bibitem{Liu2022}
Y.~Liu, D.~Li, H.~Dai, C.~Li, and R.~Zhang, ``Understanding the impact of
  environmental conditions on zero-power {Internet of Things}: An experimental
  evaluation,'' \emph{{IEEE Wireless Commun.}}, pp. 1--8, 2022.

\bibitem{tesla1904transmission}
N.~Tesla, ``The transmission of electrical energy without wires,''
  \emph{Electrical World and Engineer}, vol.~1, pp. 21--24, 1904.

\bibitem{Valenta2014}
C.~R. Valenta and G.~D. Durgin, ``Harvesting wireless power: Survey of
  energy-harvester conversion efficiency in far-field, wireless power transfer
  systems,'' \emph{IEEE Microw. Mag.}, vol.~15, no.~4, pp. 108--120, 2014.

\bibitem{Gu2021}
X.~Gu, P.~Burasa, S.~Hemour, and K.~Wu, ``Recycling ambient {RF} energy:
  Far-field wireless power transfer and harmonic backscattering,'' \emph{IEEE
  Microw. Mag.}, vol.~22, no.~9, pp. 60--78, 2021.

\bibitem{Pan2019}
G.~Pan, P.~D. Diamantoulakis, Z.~Ma, Z.~Ding, and G.~K. Karagiannidis,
  ``Simultaneous lightwave information and power transfer: Policies,
  techniques, and future directions,'' \emph{IEEE Access}, vol.~7, pp.
  28\,250--28\,257, 2019.

\bibitem{Clerckx2019}
B.~Clerckx, A.~Costanzo, A.~Georgiadis, and N.~Borges~Carvalho, ``Toward {1G}
  mobile power networks: {RF}, signal, and system designs to make smart objects
  autonomous,'' \emph{IEEE Microw. Mag.}, vol.~19, no.~6, pp. 69--82, 2018.

\bibitem{Ku2016}
M.-L. Ku, W.~Li, Y.~Chen, and K.~J. Ray~Liu, ``Advances in energy harvesting
  communications: Past, present, and future challenges,'' \emph{IEEE Commun.
  Surv. Tutor.}, vol.~18, no.~2, pp. 1384--1412, 2016.

\bibitem{Leemput2023}
D.~V. Leemput, A.~Sabovic, K.~Hammoud, J.~Famaey, S.~Pollin, and E.~D. Poorter,
  ``Energy harvesting for wireless {IoT} use cases: A generic feasibility model
  and tradeoff study,'' \emph{IEEE Internet Things J.}, pp. 1--1, 2023.

\bibitem{Costanzo2016}
A.~Costanzo and D.~Masotti, ``Smart solutions in smart spaces: Getting the most
  from far-field wireless power transfer,'' \emph{IEEE Microw. Mag.}, vol.~17,
  no.~5, pp. 30--45, 2016.

\bibitem{shen2021}
S.~Shen and B.~Clerckx, ``Joint waveform and beamforming optimization for
  {MIMO} wireless power transfer,'' \emph{{IEEE Trans.\ Commun.}}, vol.~69,
  no.~8, pp. 5441--5455, 2021.

\bibitem{Cansiz2020}
M.~Cansiz, D.~Altinel, and G.~K. Kurt, ``Effects of different modulation
  techniques on charging time in {RF} energy-harvesting system,'' \emph{IEEE
  Trans. Instrum. Meas.}, vol.~69, no.~9, pp. 6904--6911, 2020.

\bibitem{Suzhi2016}
S.~Bi, Y.~Zeng, and R.~Zhang, ``Wireless powered communication networks: an
  overview,'' \emph{IEEE Wireless Commun.}, vol.~23, no.~2, pp. 10--18, 2016.

\bibitem{Ponnimbaduge2018}
T.~D. Ponnimbaduge~Perera, D.~N.~K. Jayakody, S.~K. Sharma, S.~Chatzinotas, and
  J.~Li, ``Simultaneous wireless information and power transfer {(SWIPT)}:
  Recent advances and future challenges,'' \emph{IEEE Commun. Surveys Tuts.},
  vol.~20, no.~1, pp. 264--302, 2018.

\bibitem{Uysal2021}
M.~Uysal, S.~Ghasvarianjahromi, M.~Karbalayghareh, P.~D. Diamantoulakis, G.~K.
  Karagiannidis, and S.~M. Sait, ``{SLIPT} for underwater visible light
  communications: Performance analysis and optimization,'' \emph{{IEEE Trans.\
  Wireless Commun.}}, vol.~20, no.~10, pp. 6715--6728, 2021.

\bibitem{Liu2016}
W.~{Liu}, X.~{Zhou}, S.~{Durrani}, and P.~{Popovski}, ``{SWIPT} with practical
  modulation and {RF} energy harvesting sensitivity,'' in \emph{{Proc. IEEE
  Int. Conf. Commun. (ICC)}}, 2016, pp. 1--7.

\bibitem{Amarasuriya2016}
G.~Amarasuriya, E.~G. Larsson, and H.~V. Poor, ``Wireless information and power
  transfer in multiway massive {MIMO} relay networks,'' \emph{{IEEE Trans.\
  Wireless Commun.}}, vol.~15, no.~6, pp. 3837--3855, 2016.

\bibitem{Paolini2022}
G.~Paolini, A.~Quddious, D.~Chatzichristodoulou, D.~Masotti, S.~Nikolaou, and
  A.~Costanzo, ``An energy-autonomous {SWIPT RFID} tag for communication in the
  2.4 {GHz ISM} band,'' in \emph{3rd URSI Atlantic and Asia Pacific Radio
  Science Meeting (AT-AP-RASC)}, 2022, pp. 1--4.

\bibitem{Zhang2014}
J.~Xu, L.~Liu, and R.~Zhang, ``Multiuser {MISO} beamforming for simultaneous
  wireless information and power transfer,'' \emph{IEEE Trans. Signal
  Process.}, vol.~62, no.~18, pp. 4798--4810, 2014.

\bibitem{Veedu2022}
S.~N.~K. Veedu, M.~Mozaffari, A.~Höglund, E.~A. Yavuz, T.~Tirronen,
  J.~Bergman, and Y.-P.~E. Wang, ``Toward smaller and lower-cost {5G} devices
  with longer battery life: An overview of {3GPP} release 17 {RedCap},''
  \emph{IEEE Commun. Stand. Mag.}, vol.~6, no.~3, pp. 84--90, 2022.

\bibitem{Xu2019}
D.~Xu and H.~Zhu, ``{Secure transmission for SWIPT IoT systems with full-duplex
  IoT devices},'' \emph{IEEE Internet Things J.}, vol.~6, no.~6, pp.
  10\,915--10\,933, 2019.

\bibitem{Huang2018}
Y.~Huang, M.~Liu, and Y.~Liu, ``Energy-efficient {SWIPT} in {IoT} distributed
  antenna systems,'' \emph{IEEE Internet Things J.}, vol.~5, no.~4, pp.
  2646--2656, 2018.

\bibitem{Krikidis2019}
I.~Krikidis and C.~Psomas, ``Tone-index multisine modulation for {SWIPT},''
  \emph{IEEE Signal Process. Lett.}, vol.~26, no.~8, pp. 1252--1256, 2019.

\bibitem{Nakamoto2022}
Y.~Nakamoto, N.~Hasegawa, T.~Hirakawa, and Y.~Ohta, ``A study on {OFDM}
  modulation suitable for wireless power transfer,'' in \emph{Wireless Power
  Week (WPW)}, 2022, pp. 21--24.

\bibitem{Ilter2022}
M.~C. Ilter, R.~Wichman, M.~Säily, and J.~Hämäläinen, ``Information
  harvesting for far-field wireless power transfer,'' \emph{IEEE Internet
  Things Mag.}, vol.~5, no.~2, pp. 127--132, 2022.

\bibitem{Wang2019}
N.~Wang, P.~Wang, A.~Alipour-Fanid, L.~Jiao, and K.~Zeng, ``Physical-layer
  security of {5G} wireless networks for {IoT}: Challenges and opportunities,''
  \emph{IEEE Internet Things J.}, vol.~6, no.~5, pp. 8169--8181, 2019.

\bibitem{Goel2008}
S.~Goel and R.~Negi, ``Guaranteeing secrecy using artificial noise,''
  \emph{{IEEE Trans.\ Wireless Commun.}}, vol.~7, no.~6, pp. 2180--2189, 2008.

\bibitem{Lv2018}
L.~Lv, Z.~Ding, Q.~Ni, and J.~Chen, ``Secure {MISO-NOMA} transmission with
  artificial noise,'' \emph{{IEEE Trans. Veh. Technol.}}, vol.~67, no.~7, pp.
  6700--6705, 2018.

\bibitem{Gu2019}
Y.~Gu, Z.~Wu, Z.~Yin, and X.~Zhang, ``The secrecy capacity optimization
  artificial noise: A new type of artificial noise for secure communication in
  {MIMO} system,'' \emph{IEEE Access}, vol.~7, pp. 58\,353--58\,360, 2019.

\bibitem{Ilter2022GL}
M.~C. Ilter, E.~Basar, R.~Wichman, and J.~Hämäläinen, ``Information
  harvesting for far-field {RF} power transfer through index modulation,'' in
  \emph{{Proc. IEEE Glob. Commun. Conf.~(GLOBECOM)}}, 2022.

\bibitem{Ilter2023}
M.~C. Ilter, R.~Wichman, J.~Hämäläinen, and S.~Ikki, ``A new information
  harvesting mechanism for far-field wireless power transfer,'' in \emph{{Proc.
  IEEE Veh. Tech. Conf.~(VTC-Spring)}}, 2023, pp. 1--6.

\bibitem{Alevizos2018}
P.~N. Alevizos and A.~Bletsas, ``Sensitive and nonlinear far-field {RF} energy
  harvesting in wireless communications,'' \emph{{IEEE Trans.\ Wireless
  Commun.}}, vol.~17, no.~6, pp. 3670--3685, 2018.

\bibitem{Clerckx2022}
B.~Clerckx, J.~Kim, K.~W. Choi, and D.~I. Kim, ``Foundations of wireless
  information and power transfer: Theory, prototypes, and experiments,''
  \emph{{Proc. IEEE Proc.}}, vol. 110, no.~1, pp. 8--30, 2022.

\bibitem{Bayguzina2016}
B.~Clerckx and E.~Bayguzina, ``Waveform design for wireless power transfer,''
  \emph{IEEE Trans. Signal Process.}, vol.~64, no.~23, pp. 6313--6328, 2016.

\bibitem{Kim2020}
J.~Kim, B.~Clerckx, and P.~D. Mitcheson, ``Signal and system design for
  wireless power transfer: Prototype, experiment and validation,'' \emph{{IEEE
  Trans.\ Wireless Commun.}}, vol.~19, no.~11, pp. 7453--7469, 2020.

\bibitem{Jeganathan2008}
J.~Jeganathan, A.~Ghrayeb, and L.~Szczecinski, ``Generalized space shift keying
  modulation for {MIMO} channels,'' in \emph{{Proc. IEEE Int. Symp. on Pers.,
  Indoor and Mobile Radio Commun.~(PIMRC)}}, 2008, pp. 1--5.

\bibitem{Mesleh2015}
R.~Mesleh, S.~S. Ikki, and H.~M. Aggoune, ``Quadrature spatial modulation,''
  \emph{{IEEE Trans. Veh. Technol.}}, vol.~64, no.~6, pp. 2738--2742, 2015.

\bibitem{Poor2019}
B.~Clerckx, R.~Zhang, R.~Schober, D.~W.~K. Ng, D.~I. Kim, and H.~V. Poor,
  ``Fundamentals of wireless information and power transfer: From {RF} energy
  harvester models to signal and system designs,'' \emph{IEEE J. Sel. Areas
  Commun.}, vol.~37, no.~1, pp. 4--33, 2019.

\bibitem{DiRenzo2010}
M.~Di~Renzo and H.~Haas, ``A general framework for performance analysis of
  space shift keying {(SSK)} modulation for {MISO} correlated {Nakagami-m}
  fading channels,'' \emph{IEEE Trans. Commun.}, vol.~58, no.~9, pp.
  2590--2603, 2010.

\bibitem{Huang2022}
Z.~Huang, Z.~Gao, and L.~Sun, ``Anti-eavesdropping scheme based on quadrature
  spatial modulation,'' \emph{{IEEE Commun.\ Lett.}}, vol.~21, no.~3, pp.
  532--535, 2017.

\bibitem{yaman2015}
Y.~Wei, L.~Wang, and T.~Svensson, ``Analysis of secrecy rate against
  eavesdroppers in {MIMO} modulation systems,'' in \emph{Int. Conf. on Wireless
  Commun. Signal Process. (WCSP)}, 2015, pp. 1--5.

\bibitem{singh2021}
U.~Singh, M.~R. Bhatnagar, and T.~A. Tsiftsis, ``Secrecy analysis of {SSK}
  modulation: Adaptive antenna mapping and performance results,'' \emph{{IEEE
  Trans.\ Wireless Commun.}}, vol.~20, no.~7, pp. 4614--4630, 2021.

\bibitem{FCC}
{Federal~Commun.~Comm.}, ``{47 CFR Part15-Radio Frequency Devices},''
  Washington, DC, USA, 2016.

\bibitem{Ho2012}
X.~Zhou, R.~Zhang, and C.~K. Ho, ``Wireless information and power transfer:
  Architecture design and rate-energy tradeoff,'' \emph{IEEE Trans. Commun.},
  vol.~61, no.~11, pp. 4754--4767, 2013.

\bibitem{Basar2017}
E.~Basar, M.~Wen, R.~Mesleh, M.~Di~Renzo, Y.~Xiao, and H.~Haas, ``Index
  modulation techniques for next-generation wireless networks,'' \emph{IEEE
  Access}, vol.~5, pp. 16\,693--16\,746, 2017.

\end{thebibliography}

\end{document}